\documentclass[useAMS,usenatbib]{mn2e}
\usepackage[pdfpagemode={UseOutlines},bookmarks,bookmarksopen,colorlinks,citecolor={blue},urlcolor={red}]{hyperref}
\usepackage[flushleft]{threeparttable}
\usepackage{booktabs,caption}
\usepackage{graphicx}
\usepackage{amsmath}
\usepackage{bm}

\usepackage{subfigure}
\usepackage{amssymb}
\usepackage{tabularx}
\usepackage{natbib}
\usepackage{float}
\usepackage{etoolbox}

\usepackage{natbib}
\usepackage[usenames,dvipsnames]{xcolor}

\usepackage[normalem]{ulem}

\newcommand{\bea}{\begin{eqnarray}}
\newcommand{\eea}{\end{eqnarray}}

\newcommand{\Ms}{~{\rm M}_\odot}
\newcommand{\flux}{~{\rm GeV} ~{\rm cm}^{-2} ~{\rm s}^{-1}}
\newcommand{\nc}{{\rm NSC}}
\newcommand{\bh}{{\rm BH}}
\newcommand{\msp}{{\rm MSP}}
\newcommand{\cv}{{\rm CV}}

\definecolor{darkgreen}{rgb}{0, .5, 0}

\makeatletter

\patchcmd{\NAT@citex}
  {\@citea\NAT@hyper@{%
     \NAT@nmfmt{\NAT@nm}%
     \hyper@natlinkbreak{\NAT@aysep\NAT@spacechar}{\@citeb\@extra@b@citeb}%
     \NAT@date}}
  {\@citea\NAT@nmfmt{\NAT@nm}%
   \NAT@aysep\NAT@spacechar\NAT@hyper@{\NAT@date}}{}{}

\patchcmd{\NAT@citex}
  {\@citea\NAT@hyper@{%
     \NAT@nmfmt{\NAT@nm}%
     \hyper@natlinkbreak{\NAT@spacechar\NAT@@open\if*#1*\else#1\NAT@spacechar\fi}%
       {\@citeb\@extra@b@citeb}%
     \NAT@date}}
  {\@citea\NAT@nmfmt{\NAT@nm}%
   \NAT@spacechar\NAT@@open\if*#1*\else#1\NAT@spacechar\fi\NAT@hyper@{\NAT@date}}
  {}{}

\makeatother

\newcommand{\gae}{\lower 2pt \hbox{$\, \buildrel {\scriptstyle >}\over {\scriptstyle \sim}\,$}} 

\newcommand{\lae}{\lower 2pt \hbox{$\, \buildrel {\scriptstyle <}\over {\scriptstyle \sim}\,$}} 

\title[Gamma-ray and X-ray emission from the Galactic centre]{Gamma-ray and X-ray emission from the Galactic centre: hints on the nuclear star cluster formation history}

\author[Arca-Sedda, M. and Kocsis, B. and Brandt, T.~D.]{
Manuel ~Arca-Sedda$^{1}$\thanks{e-mail: m.arcasedda@ari.uni-heidelberg.de} and Bence ~Kocsis$^{2}$ and Timothy D. ~Brandt$^{3,4}$ \\
$^{1}$Zentrum f\"{u}r Astronomie der Universit\"{a}t Heidelberg, Astronomisches Rechen-Institut, M\"onchhofstr. 12-14, 69120 Heidelberg\\
$^{2}$Institute of Physics, E\"otv\"os University, P\'azm\'any P. s. 1/A, Budapest, 1117, Hungary\\
$^{3}$Institute for Advanced Study, Einstein Dr., Princeton, NJ, 08540 USA\\
$^{4}$University of California, Santa Barbara, Santa Barbara, CA, 93106, USA
}

\begin{document}
\date{Revised to }

\pagerange{\pageref{firstpage}--\pageref{lastpage}} \pubyear{2015}

\maketitle

\label{firstpage}

\maketitle

\begin{abstract}
The Milky Way centre exhibits an intense flux in the gamma and X-ray bands, whose origin is partly ascribed to the possible presence of a large population of millisecond pulsars (MSPs) and cataclysmic variables (CVs), respectively.
However, the number of sources required to generate such an excess is much larger than what is expected from in situ star formation and evolution, opening a series of questions about the formation history of the Galactic nucleus. In this paper we make use of direct $N$-body simulations to investigate whether these sources could have been brought to the Galactic centre by a population of star clusters that underwent orbital decay and formed the Galactic nuclear star cluster (NSC). Our results suggest that the gamma ray emission is compatible with a population of MSPs that were mass segregated in their parent clusters, while the X-ray emission is consistent with a population of CVs born via dynamical interactions in dense star clusters. Combining observations with our modelling, we explore how the observed $\gamma$ ray flux can be related to different NSC formation scenarios.
Finally, we show that the high-energy emission coming from the galactic central regions can be used to detect black holes heavier than $10^5\Ms$ in nearby dwarf galaxies.
\end{abstract}


\section{Introduction}

The {\it Fermi} satellite's discovery of strong excess emission in the Milky Way Galaxy centre opened a series of questions about the physical origin of this phenomenon. The excess is characterised by a nearly spherical morphology, and extends up to $\sim 1 - 3$ kpc from the Galaxy's supermassive black hole (SMBH). One possible explanation for such a strong signal is the annihilation of $\sim 30$ GeV dark matter particles \citep{hooper11}. In this case, the Galactic centre would provide the first evidence of dark matter particles beyond the Standard Model interacting with the electromagnetic sector. However, such exotic explanations presume that astrophysical processes cannot account for the observed emission. Possible alternatives to dark matter annihilation, among others, are millisecond pulsars (MSPs), rapidly rotating neutron stars observed throughout the galaxy and characterised by a gamma ray spectrum similar to that observed for the excess \citep{Abazajian11}; highly magnetized young pulsars born in the star-forming nuclear star cluster (NSC) \citep{Leary15,oleary16}; injection of cosmic-ray protons \citep{carlson14}; or cosmic ray outbursts \citep{petrovic14}. 
While a diffuse background would be expected from annihilation in a smooth dark matter profile \citep{bartels16,mishra17}, observations instead indicate an unresolved population of gamma ray point sources, consistent with the hypothesis that MSPs play a significant role in the development of gamma ray emission \citep{daylan16,fermi17}.

{\it Fermi}'s resolution is insufficient to study the morphology of the excess over the inner few pc. The excess is consistent with a compact, unresolved ($\lesssim$10~pc) source or set of sources, plus much more extended emission \citep{Abazajian14,brandt15}, or with a steep cusp toward Sgr A* \citep{daylan16}. The morphology of this inner region is accessible in X-rays.
Recent observations with the NuSTAR satellite show a complex X-ray structure in the inner 10 pc from Sgr A*, characterised by a nearly spherical structure and emitting filaments. The source of emission may be an unresolved population of cataclysmic variables (CVs, see \cite{mukai17} for a review on the argument)  with mass $\sim 0.9\Ms$ \citep{mori15}, a population of MSPs \citep{perez15}, or a more heterogeneous population of MSPs, CVs and X-ray binaries.
\cite{hailey16} propose that the emission is likely due to intermediate polars (IPs), a type of CV with longer orbital periods and non-synchronized orbits compared to polars \citep{patterson94,pretorius13}. The authors note that the central X-ray emission profile is quite similar to the luminosity profile of the Galactic NSC, thus suggesting a stellar origin for the X-rays. 
Understanding the nature of the gamma-ray excess and its possible connection with the X-ray excess may shed light on the extreme processes that take place in the vicinity of an SMBH.

In this paper, we use direct $N$-body simulations to examine the role of infalling globular clusters (GCs) in shaping the observed gamma-ray and hard X-ray profiles. GCs efficiently form MSPs, CVs, and X-ray binaries due to the high likelihood of close dynamical encounters. Using semianalytic arguments and comparisons to extant GCs, \citet{brandt15} showed that the observed gamma-ray flux is consistent with the emission of MSPs that were delivered to the Galactic centre by inspiralling GCs. Similarly, we point out that infalling clusters would also deposit their CV populations around the Galactic centre. Indeed, CVs are expected to form via dynamical encounters in dense stellar environments \citep{ivanova06,belloni16,dieball17}, and their lifetimes are estimated to lie in the range $10^8$-$10^{11}$ yr \citep{kolb96,mukai17}. CVs with period shorter than 2.2 hr have a lifetime $>10^9$ yr\footnote{i.e. those below the so-called ``period gap'', and constitute $\sim 30\%$ of the CV population \citep{kolb96}, although it is difficult to determine the actual fraction due to observational selection effects \citep{mukai17} or to the complex modelling required \citep{podsiadlowski03,goliasch15}.}.
Recent observations in the far UV have outlined the presence of a population of both CVs and MSPs mass-segregated into the core of the NGC 6397 GC \citep{dieball17}. This poses interesting questions about the formation and evolution of such objects, confirming at the same time their presence in the inner regions of GCs. 

Dynamically formed MSPs and CVs carry information on the GC infall history of the Galactic center. Using our simulations, we examine the formation of the central Milky Way nuclear star cluster, and predict the distribution of MSPs and CVs. We use this information to investigate the implications for gamma ray and hard X-ray emission profiles. 

\subsection{Controversy of the MSP origin of the Fermi excess}

In the MSP interpretation of the Fermi excess, the observed gamma ray flux requires at least $10^3$ MSPs \citep{bednarek13}, a number that seems exceedingly large with respect to our current knowledge of in situ MSP formation mechanisms in the Galactic centre. The MSPs could have formed in a dense
stellar environment, such as a GC, and have been delivered to the central region by the inspiral of dense systems \citep{brandt15}. Indeed, known as ``recycled pulsars'', MSPs form primarily in binary systems. The high stellar encounter rates in dense systems, facilitates to decrease the binary separation, until the neutron star's companion transfers material and angular momentum, reducing the neutron star's magnetic field and increasing its spin rate \citep{michel87,Bhattacharya91}. During this phase, lasting $\sim 10^7 - 10^9$ yr, the system is observable as a low mass X-ray binary \citep[LMXB,][]{ivanova08}. After the mass transfer stops, the MSP phase will live beyond $\sim 10^{10}$ yr. 

Given their long life time, MSPs constitute a promising source for the Galactic $\gamma$ ray flux, although many objections have been put forth against this scenario.
\cite{haggard17} argued that if the observed gamma-ray flux is only due to MSPs, they should have observed $\sim$1000 LMXBs within $10 {\rm deg}$ from the Galactic centre ($\sim 1.4$ kpc @ $8$ kpc), but only $\sim 40-80$ LMXBs have been observed there with INTEGRAL \citep{lavigne98}. The MSP scenario thus requires that most LMXB activity, and hence MSP creation, ceased long ago. If the MSP population is old, the present-day emission depends strongly on the gamma-ray efficiency, which varies as a function of spin-down power \citep{oleary16}. The observed excess is inconsistent with MSPs assuming a constant gamma-ray efficiency \citep{hooper16}. However, recently \citet{fragione17} showed that the observed emission is consistent with the expected MSP emission accounting for spin down effects. Finally, \cite{hooper16} claimed that the MSP luminosity function requires several very bright (and individually resolvable) MSPs around the Galactic center in order to explain the excess. 

With the ongoing development of hybrid strategies to infer the actual number of MSPs from {\it Fermi} data \citep{bartels16,bhakta17} and with future radio observations with the square kilometer array (SKA) it may be possible to detect MSPs directly and map out their distribution in the MW inner 10 pc \citep{dewdeney09,macquart15,brandt15,calore16}, as recently suggested by \cite{abbate17}. 

\subsection{Formation of the nuclear star cluster}
The distribution of MSPs and CVs in the Galactic center carries important information about the formation history of the nuclear star cluster (NSC). The NSC is a massive star cluster surrounding Sgr A* which is characterised by a very compact size (half-mass radius $\sim$4.2 pc) and a total mass of $2.5\times 10^7\Ms$ \citep{schodel14,gallegocano2017,schodel2017}. In what is called the ``dry-merger'' scenario, NSCs are assembled by the sequence of mergers of dense star clusters that spiral toward the galaxy centre due to dynamical friction \citep{Trem75,Trem76a,Trem76,Dolc93,AMB,antonini13,gnedin14,ASCD14a,ASCD14b,ASCD15He,ASCD16a,ASCD16b}. 
During this process, GCs bring their stellar content to the Galactic centre
altering the stellar population therein \citep{antonini14,ASCD16b}.
The ``dry'' scenario contrasts with a ``wet'' formation in which gas was brought to the Galactic centre where stars formed in-situ \citep{Mil04,nayakshin,antonini13,aharon15}.

\bigskip
In this paper we use state-of-the-art $N$-body simulations to investigate whether a ``dry'' origin of the Galaxy's nuclear star cluster (NSC) can account for the Galactic centre's gamma-ray and X-ray emission observed by Fermi and NuSTAR, and examine implications for MSP and CV source candidates.\footnote{After the original submission of this paper, a paper appeared on the predicted gamma-ray emission of the Galactic bulge for a delivered population of MSPs accounting for the spin-down effect \citep{fragione17}. Their calculation leads to a $\gamma$ flux an order of magnitude larger than observed in the case in which the MSP spindown effect is taken into account. } We also  determine the number of stellar mass BHs delivered to this region by GCs.

The paper is organized as follows: in Section \ref{model} we describe our numerical setup and our models; in Section \ref{results} we discuss the outcomes of the simulations and in Section \ref{end} we draw the conclusions of this work.

\section{Numerical Model}
\label{model}

In order to investigate how the GCs' initial conditions can affect the gamma-ray emission profile, we used the set of direct $N$-body simulations presented in \citet{ASCD15He}, which was used to model the long-term evolution of the galaxy Henize 2-10. This dwarf starburst galaxy is an excellent observational target to study NSC formation, as it contains 11 massive star clusters with masses in the range $(0.2-2)\times 10^6\Ms$ \citep{ngu14}, orbiting at distances $\lesssim 200$ pc around an SMBH with mass $2.6\times 10^6\Ms$ \citep{reines}. 
In the rest of the paper we do not distinguish young massive star clusters and globular clusters and refer to both as GCs.
These $N$-body simulations confirmed that the star clusters likely segregate to the Galactic centre rapidly on very short timescales, $\sim$100 Myr, leading to the formation of a dense NSC with mass $\simeq (4-6) \times 10^6\Ms$ depending on the star clusters' initial conditions (see Table \ref{nsc}) \citep{ASCD15He}. We list the masses of the assumed GCs in Table~\ref{ICs}.

\begin{table}
\caption{Star clusters properties and mass deposited}
\begin{center}
\begin{tabular}{ccccc}
\hline   
ID & $M_{\rm GC}$ &  \multicolumn{3}{c}{$ M_{\rm GC}(<10~{\rm pc})/M_{\rm GC}~~(\%)$} \\
            & $10^6 \Ms $ &  { S1} & {  S2} &{  S3}\\
\hline
\hline
   $1$  &$2.29$ &$54.6$&$54.6$&$66.4$\\
   $2$  &$0.92$ &$42.8$&$44.8$&$42.9$\\
   $3$  &$1.14$ &$15.3$&$38.6$&$31.3$\\
   $4$  &$0.91$ &$47.4$&$55.3$&$79.2$\\
   $5$  &$0.40$ &$0.02$&$33.4$&$0$\\
   $6$  &$0.40$ &$0$   &$37.9$&$0$\\
   $7$  &$0.46$ &$0$   &$0$   &$0$\\
   $8$  &$0.45$ &$1.1$ &$0$   &$1.0$\\
   $9$  &$0.20$ &$0$   &$40.8$&$0$\\
   $10$ &$0.45$ &$0$   &$57.1$&$0$\\
   $11$ &$0.20$ &$0$   &$41.8$&$0$\\
\hline 
\end{tabular} 
\end{center}
\begin{tablenotes}
\item 
Col. 1: GC identification number. Col. 2: GC mass. Col 3-6: percentage of the GC mass deposited within 10 pc from the SMBH for models S1, S2, and S3. 
\end{tablenotes}
\label{ICs}
\end{table}

This model represents the inner regions of a galaxy much smaller than the Milky Way, but provides a view on its galactic nucleus before the formation of its NSC. 
Indeed, while we know the current morphology and mass distribution of the Galactic NSC, the larger-scale properties of the Milky Way's nucleus (and its properties before NSC formation) remain uncertain.  

The formation of an NSC leads to a significant enhancement of the central density slope \citep{merritt06,AMB,ASCD15He}. Before this took place, the stellar distribution in the Galactic centre could have been different from today. The minimum negative radial density exponent required to achieve a self-consistent distribution around a SMBH is $\gamma=0.5$, which is the choice assumed here \citep{merritt06,Merri13}. 

Our galaxy model is a truncated Dehnen sphere \citep{Deh93,ASCD15He}:
\begin{equation}
\rho_{\rm D}(r)=\frac{(3-\gamma)M_g}{4\pi r_g^3}\left(\frac{r}{r_g}\right)^{-\gamma}\left(1+\frac{r}{r_g}\right)^{-4+\gamma}\frac{1}{{\rm cosh}(r/r_{\rm cut})} ,
\label{denh}
\end{equation}
with $\gamma = 0.5$ the negative inner density slope, $M_g=1.6\times 10^9\Ms$ the total galaxy mass and $r_g=110$ pc its length scale. The truncation radius $r_{\rm cut}=150$ pc allowed us to model self-consistently the inner region of the Galaxy. 

This choice of parameters is chosen to roughly represent the observed cumulative mass profile and velocity dispersion (see \citealt{ASCD15He} for more details).
The outer cut at $150$ pc is set by computational limitations while keeping the necessary high resolution of the inner regions. This approach allows us to create a self-consistent model that reproduces the whole region inside $r_{\rm cut}$, whose evolution is driven by two-body relaxation processes. In other words, cutting the density profile with the exponential cut in Equation \eqref{dens} ensures a correct representation of the dynamics inside $r_{\rm cut}$ avoiding spurious relaxation processes related to the absence of particles outside of this region.

The central SMBH is modelled with a point-like particle with mass $M_\bh = 2.6\times 10^6 \Ms$, in agreement with the observational estimates in Henize 2-10 \citep{reines,reines12,reines16}.
Note that this value is quite similar to the Sgr A* mass, $3.6^{+0.2}_{-0.4} \times 10^6\Ms$ \citep{ghez08,gillessen09,schodel09}, making the Henize 2-10 nucleus models capable of testing the dry-merger origin for the Milky Way's NSC.

We analysed three different initial conditions for the geometry of the distribution of the globular cluster and the presence of the SMBH: 
\begin{enumerate}
\item model S1, in which the number density distribution of clusters were assumed to follow the distribution of the background galaxy, 
\item model S2, corresponding to the assumption that the clusters are all initially located in the same plane and co-rotate, and
\item model S3, where clusters have the same initial conditions as in model S1, but in this case the galaxy does not contain any central SMBH. 
\end{enumerate}

Models S1 and S2 represent two limiting cases. While in S1 the star clusters are distributed in phase space according to the Galactic background, in S2 they are distributed over the same plane and have distances from the central SMBH smaller than 150 pc. The recent finding of a significant rotation in the Galactic nuclear cluster \citep{feldmeier14} and its possible connection to the dry-merger scenario \citep{tsatsi17} indicates that the star clusters that contributed to its assembly likely included initial orbital properties in between models S1 and S2. 

 The final NSC angular momentum is a fraction of the sum of all the merging GCs' angular momenta, which are partly erased by the action of dynamical friction. For initially co-planar orbits, GCs momenta are parallel and have the same sign, thus leading to a rotating NSC. 
If GCs move on random orientation orbits, the mergers lead to a partial cancellation of the angular momentum and produce a slowly rotating NSC.\footnote{Clearly, an exactly null angular momentum can be achieved only in special configuration.} 

The recent discovery of a population of massive young clusters orbiting in a disc configuration \citep{ngu14} also motivates the S2 model, which allows us to investigate the possibility of NSC formation by GCs formed in the Galactic disk.
Finally, comparing models S1 and S3 allows us to highlight the role played by the central SMBH in shaping the nuclear cluster's properties and its stellar content. 
We refer the reader to \cite{ASCD15He} for further details on the GCs' initial conditions and orbital properties.

In order to obtain a reasonable balance between computational load and the resolution of our cluster models, we allowed a difference between the mass of cluster stars, $m_c$, and that of galaxy stars, $m_g$, assuming $m_g/m_c=8$. We ran several tests in order to ensure that the results are not affected by such a choice.

The simulations have been carried out using the \texttt{HiGPUs} code \citep{Spera}, a 6th order Hermite integrator with block time-steps that runs on hybrid GPU-CPU platforms. 
In all the simulations, we set a softening parameter $\epsilon = 0.02$ pc in order to smooth strong gravitational interactions.

\subsection{Scaling strategy}
\subsubsection{Adapting the $N$-body model to the Milky Way nucleus}
\label{scale}

Our simulations were originally tailored to the Henize 2-10 galaxy, and followed the evolution of 11 young massive clusters with masses in the range $(0.2-2.6)\times 10^6\Ms$ \citep{ngu14}, orbiting around an SMBH with mass $M_\bh = 2.6\times 10^6\Ms$ \citep{reines,reines16}.
Due to its nature, $N$-body modelling can be easily adapted to different systems by simply rescaling the particle mass and positions, and by changing the velocity- and time-scales accordingly.
In order to rescale the simulation results to the Milky Way, we rescale the masses and positions to match the observed total mass and effective radius 
\begin{align}\label{eq:rescale1}
m_i &\rightarrow \frac{M_{\rm MWNSC}}{M_\nc} \times m_i\,, \\
\bm{r}_i &\rightarrow \frac{r_{\rm MWNSC}}{r_\nc} \times \bm{r}_i\,. \label{eq:rescale2}
\end{align}
 According to the observed properties of the Galactic NSC, we assumed $M_{\rm MWNSC} = 2.5\times 10^7 \Ms$ and $r_{\rm MWNSC} = 4.2$ pc \citep{schodel14,schodel2017}.

As dynamical friction drags the star clusters toward the Galactic centre, they lose stars from their outskirts due to tidally-enhanced evaporation. Table \ref{ICs} shows the fraction of GC mass deposited within the inner $10$ pc for the three models investigated. 
According to Table \ref{ICs}, we find that $\sim 20-30\%$ of the mass of the most massive clusters (1-4) is deposited into the NSC. 

After the NSC formation, nearly $60-70\%$ of the total GC mass is deposited in the innermost $20$ pc, $20\%$ is dispersed between 20-60 pc and the remaining mass is dispersed around the Galactic centre, up to 100 pc and beyond. Since the numerical costs limited our simulations to within the innermost 200 pc region of the Galactic bulge, we cannot model the mass deposited from massive GCs falling in from larger distances. However, our simulations indicate that
a significant fraction of the emission produced by MSPs comes from 
an extended region in the Galactic bulge, rather then from the central NSC, in agreement with observations \citep{Abazajian14,daylan16}.
We find that a well detectable central NSC forms in all of the simulations within $\sim 15-80$ Myr mainly due to the merger of the four most massive clusters. The NSC mass and sizes are summarized in Table \ref{nsc}.

\begin{table}
\caption{Masses and sizes of the NSCs formed in the simulations.}
\begin{center}
\begin{tabular}{ccc}
\hline 
model & $M_\nc$   & $r_\nc$  \\ 
      & $10^6\Ms$ & pc       \\
\hline
S1    &    4.7      &  4.2   \\ 
S2    &    6.0      &  2.6   \\
S3    &    5.1      &  2.0   \\
\hline
\end{tabular}
\end{center}
\begin{tablenotes}
\item Col. 1: model name. Col. 2: NSC mass. Col. 3: NSC effective radius. 
\end{tablenotes}
\label{nsc}
\end{table}

The newly-born NSC contains both ``cluster stars,'' dragged into the galactic centre from the infalling clusters, and ``galaxy stars,'' which were already present in the inner galaxy and remained trapped within the NSC during its assembly.

We stress that this definition of ``galaxy stars'' differs from the typical observer's definition when studying the NSC.  Most observers use ``galaxy stars'' to refer to interlopers: stars whose projected positions place them in the NSC, but that in reality are foreground stars very far away from the inner few pc.  Our use of the term refers to bona fide members of the NSC that were not initally members of one of the infalling GCs.  By tagging each star, we can trace its full dynamical history and attribute it to a source population.

Our definition of the NSC's size and mass follows \cite{ASCD15He}; we select as NSC members all stars moving inside the ``bump'' observed in the surface density profile.  This is simpler than the typical observational approach.  To extract the NSC's properties, observers usually fit the observed surface brightness with a combination of known profiles, like S\'ersic and Core-S\'ersic \citep{cote06,Turetal12}, and use these to infer the mass and density profile.  In our case the surface density bump provides a rough estimate of NSC size, but the boundary of the NSC (which is really just a power-law distribution of stellar density) remains somewhat fuzzy, and the fraction of NSC members that originated in GCs depends on this boundary.

For instance, in simulation S1 the NSC, as observed in the global surface density profile, consists of a clear overdensity extending up to 10 pc characterized by an effective radius $R_e\simeq 4.2$ pc. The mass enclosed within a 10 pc radius around the SMBH is $4.7\times 10^6\Ms$, as shown in Table \ref{nsc}, but
the GC debris mass deposited inside of 10 pc is $\simeq 2.3\times 10^6\Ms$. Therefore, in this case the orbitally segregated GCs represent $48\%$ of the total NSC mass. The other 52\% of the mass is ``galaxy stars'' in our terminology: bona fide NSC members that were born outside of GCs.

\begin{figure}
\centering
\includegraphics[width=8cm]{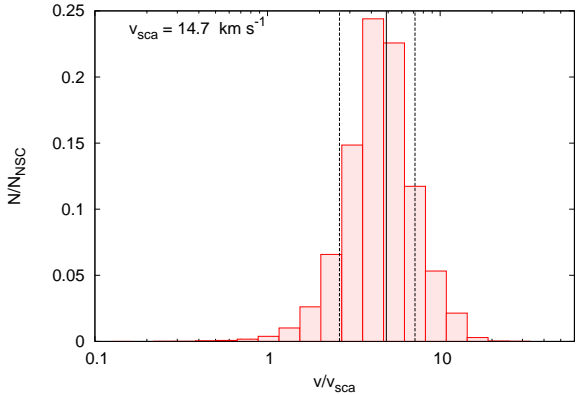}
\caption{ Velocity distribution of stars orbiting within 10 pc. The vertical line represent the mean (straight black line) and the boundaries delimited adding to this value $\pm 2\sigma$ (dotted black lines).}
\label{distriV}
\end{figure}

We can verify whether our ``galaxy stars'' and former GC stars are bound to the NSC  by calculating the velocity distribution within the NSC radius ($r\lesssim 10$ pc), and labeling as ``contaminants'' the stars having a velocity larger than a given threshold.  We adopt as our contaminants stars with $v>2\sigma$, where $\sigma$ is the velocity dispersion; at these velocities the stars can travel far from the NSC.  Figure \ref{distriV} shows that only $15\%$ of the ``galaxy stars'' orbiting in the inner $10$ pc have a velocity larger than the threshold.  The rest are bona fide NSC members.

These simulations confirmed that a dense NSC can form on a very short timescale from clusters that sink in from the inner $\lesssim 200$ pc of the Galactic bulge.

Next, we discuss the strengths and limitations of the rescaling procedure.

\subsubsection{Surface density and velocity dispersion}

\begin{figure*}
\centering
\includegraphics[width=8cm]{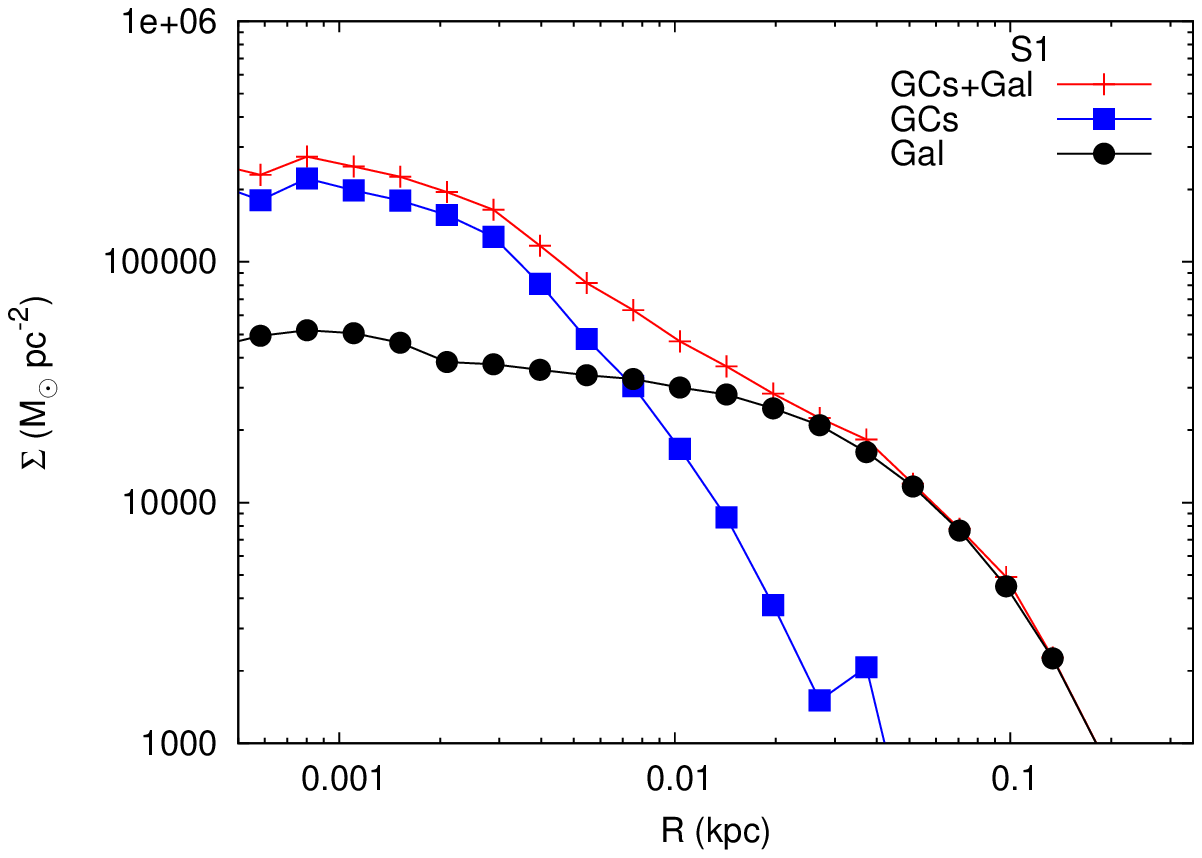}
\includegraphics[width=8cm]{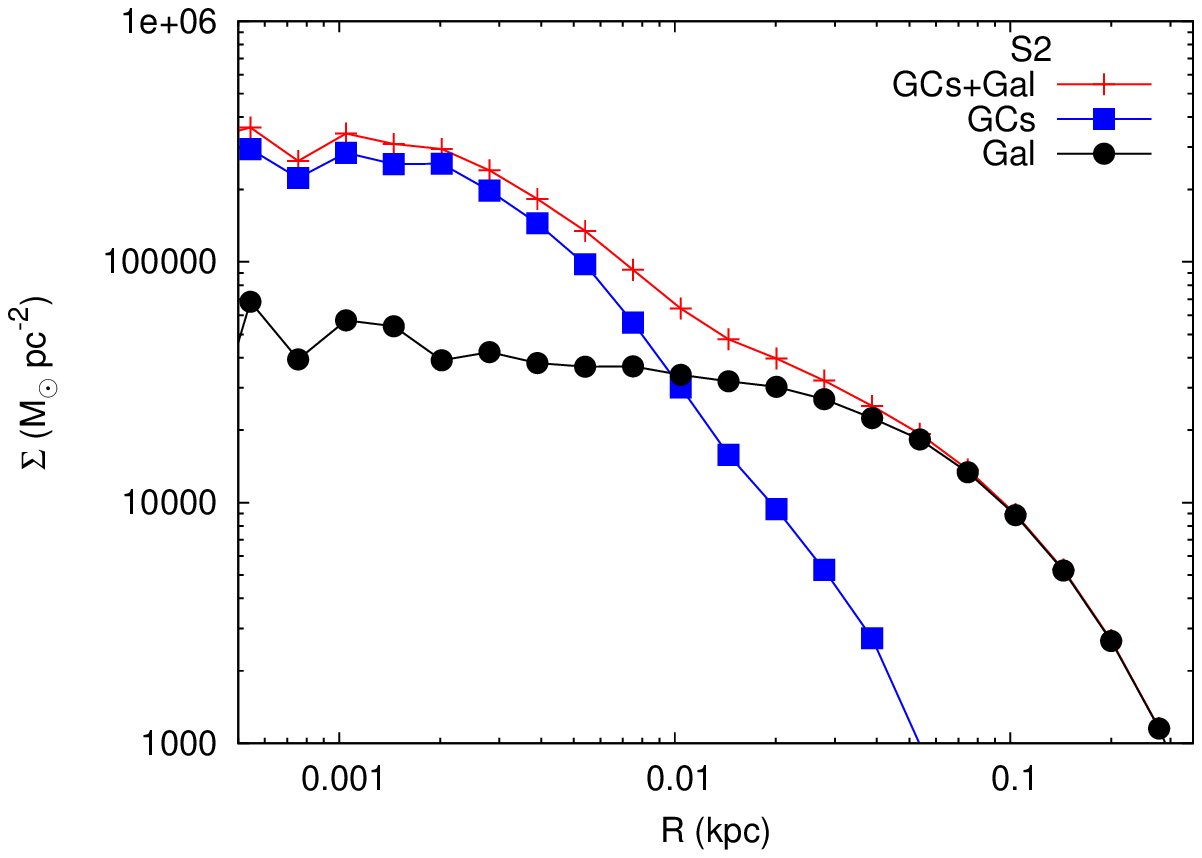}\\
\includegraphics[width=8cm]{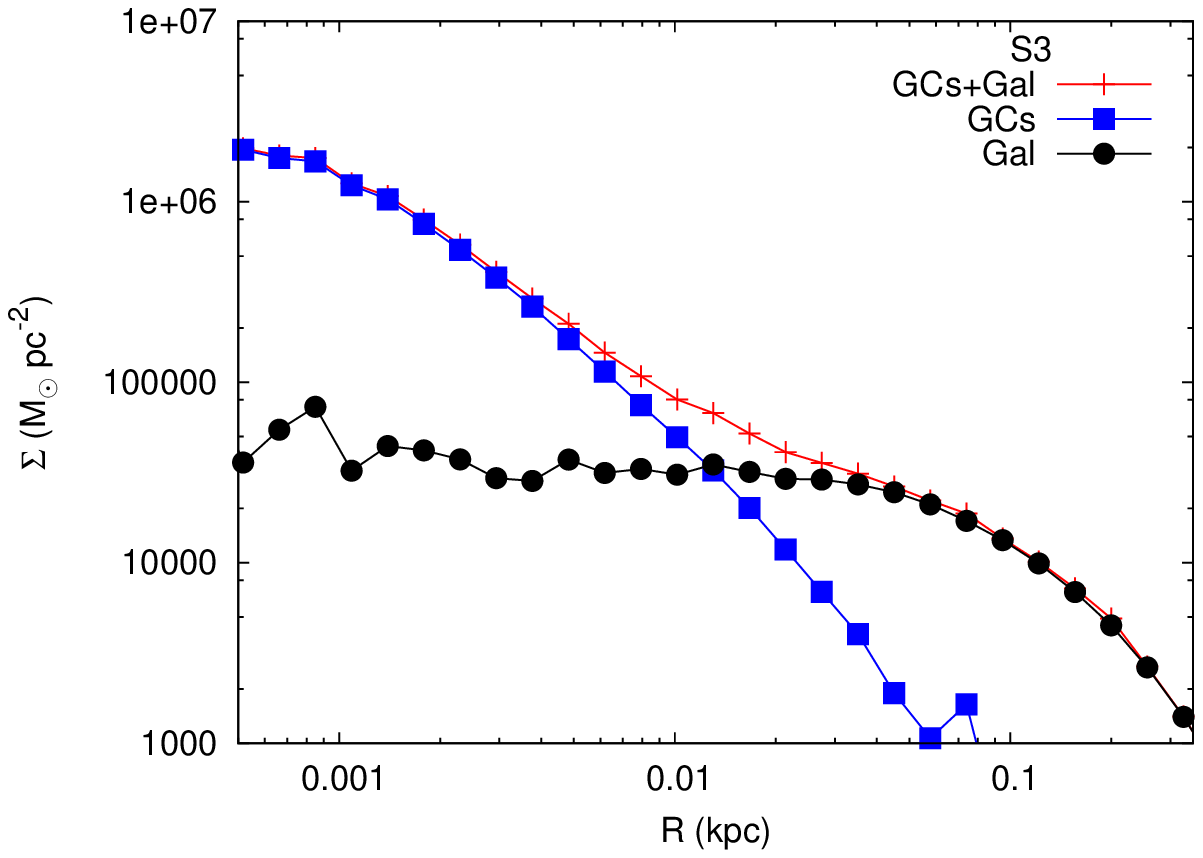}
\includegraphics[width=8cm]{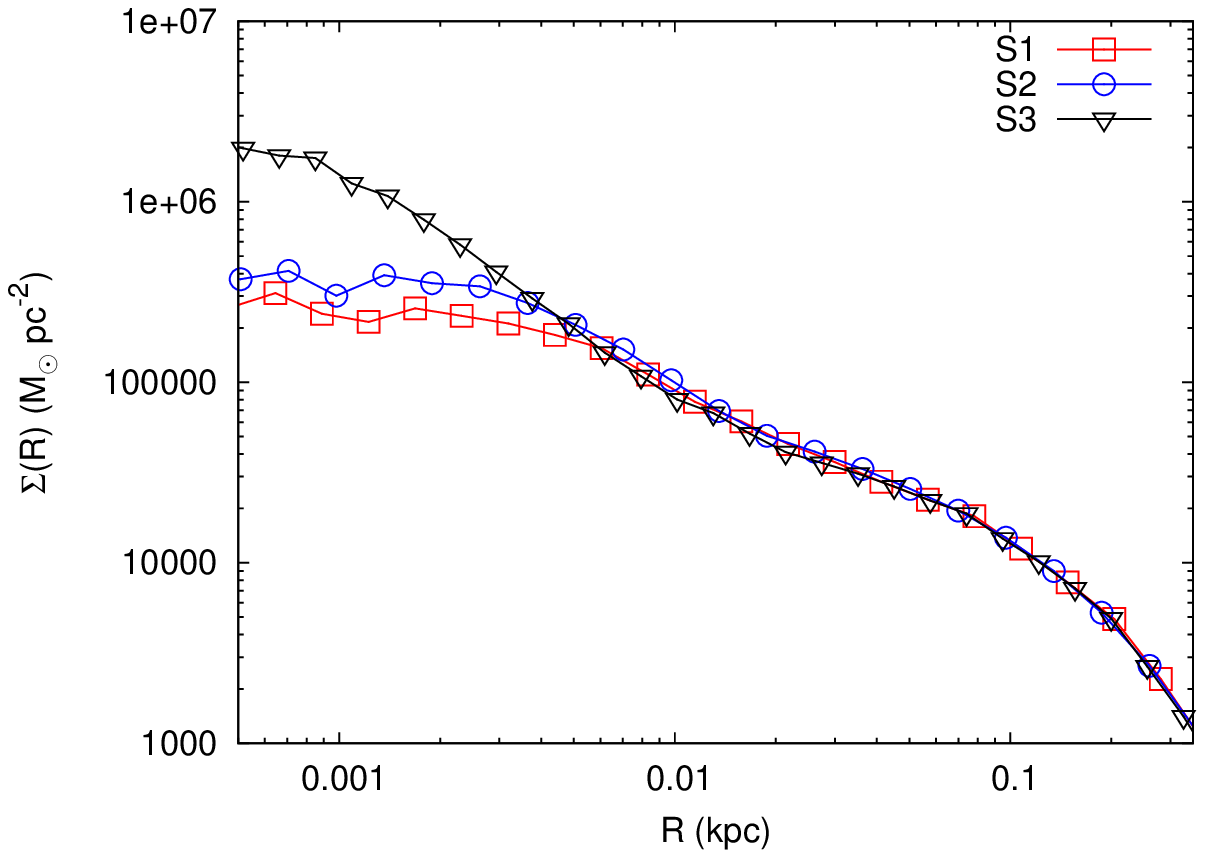}
\caption{Present day surface density profile for different initial conditions shown in Table~\ref{params}: spherical GC distribution (top left panel, model S1), disk like distribution of GCs (top right panel, model S2), and spherical GC distribution without a SMBH (bottom left panel, model S3). The solid red, blue dashed, and black dotted lines represent the total profile,  the star clusters', and the background galaxy's contributions, respectively. The bottom right panel represents a comparison between the overall surface densities for the three models investigated.}
\label{dens}
\end{figure*}

Figure \ref{dens} shows the contribution of star clusters and background to the total surface density profile of the Galactic Centre in our rescaled simulations. It is evident that in all the cases the GCs' debris dominates over the initial galaxy density in the inner 10 pc. The NSC component can be well described by a simple relation
\begin{equation}
\Sigma(R)=\Sigma_0\left[ 1 + \left(\frac{R}{R_0}\right)  \right]^{-\zeta}.
\end{equation}
The best fit parameters, listed in Table \ref{params}, of models S1 and S2 in which an SMBH is present, are in remarkably good agreement with earlier numerical calculations tailored to the Milky Way nucleus \citep{AMB}. However, our calculations are characterised by a slightly larger central surface density, due to the fact that we are rescaling our models to an NSC 1.5 times heavier than in \citet{AMB}.
Assuming that the mass and luminosity profiles follow the same behaviour, we found that our $\zeta$ are compatible with the best-fitting observational estimates, which lie in the range $0.3-0.8$ \citep{schodel14}. We find further agreement with \cite{AMB} in terms of the 3D density profile. For instance, using a modified Hubble law \citep{rood72,AMB}, 
\begin{equation}
\rho(r) = \rho_0\left[ 1+\left(\frac{r}{r_0}\right)^2\right]^{-1.5} ,
\end{equation} 
the best fit parameters in model S1 are $\rho_0 = (8.2\pm0.4)\times 10^4 \Ms$ pc$^{-3}$ and $r_0=3.3\pm 0.4$ pc.

\begin{table}
\begin{center}
\caption{NSC surface density parameters}
\label{params}
\begin{tabular}{cccc}
\hline
model & $\Sigma_0$   & $R_0$ &$ \zeta$  \\ 
      & $\Ms$ pc$^{-2}$ & pc &       \\
\hline
S1 & $ 0.28\pm 0.01$ & $ 2.1\pm0.3$& $ 0.59\pm0.10$\\
S2 & $ 0.34\pm 0.01$ & $ 4.1\pm1.1$& $ 0.88\pm0.26$\\
S3 & $ 2.18\pm 0.08$ & $ 1.2\pm0.1$& $ 0.82\pm0.06$\\
\hline
\end{tabular}
\end{center}
\begin{tablenotes}
\item Col. 1: model name. Col. 2: central surface density. Col. 3: scale radius. Col. 4: surface density slope.
\end{tablenotes}
\end{table}

The kinematics of the cluster are also in agreement with observational and numerical estimates, as our NSC models are characterized by a radial velocity dispersion of $\sim 100$ km s$^{-1}$ at 1 pc from the SMBH.

\begin{figure}
\includegraphics[width=8cm]{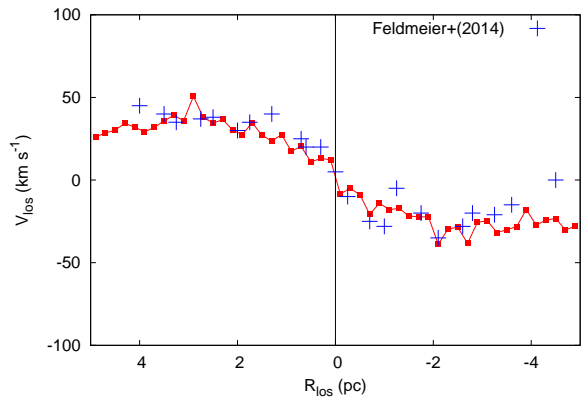}
\caption{ Line-of-sight velocity profile in our model S1 (red filled squares) compared with observed values provided by \citet{feldmeier14} (blue crosses).}
\label{los}
\end{figure}

For instance, Figure \ref{los} shows the line-of-sight (LOS) velocity profile for our model S1, rescaled to the MW nucleus. 
The profile that we show is the average of the profiles computed along the $x$, $y$, and $z$ reference axes in our model (the actual observed LOS velocity profile depends on the location of an observer in the Galaxy).  More refined methods of inferring the LOS velocity profile exist, for example the kinemetric approach \citep{krainovic06}.  However, our calculations of the high-energy emission from the inner Galaxy depend only weakly on the NSC kinematics, and our averaged LOS velocity profile agrees well with the observational results of \cite{feldmeier14} (c.f. their Figure 13). 

\subsubsection{Relaxation timescale}
Another important parameter to consider in converting the Henize 2-10 galaxy simulations to a Milky Way model is the time scale. The long-term evolution of gravitating systems is generally driven by two-body encounters over a relaxation time-scale (\citealt{spitzer87})
\begin{equation}
t_r = \frac{0.33\sigma^3}{G^2\rho {m_{\rm eff}} \ln\Lambda},
\label{relax}
\end{equation}
where $\sigma$ is the one-dimensional velocity dispersion, $\rho$ is the mean stellar density, 
$m_{\rm eff}=\langle m^2\rangle/\langle m\rangle$ is the so-called effective mass, defined as the ratio of the mean-squared stellar mass to the mean stellar mass,

and $\ln\Lambda$ is the Coulomb logarithm.
The SMBH's gravity dominates the dynamics within its influence radius $r_{\rm infl} = GM_\bh/\sigma^2$. Within this region, $\ln\Lambda = \ln(r_{\rm infl}\sigma^2/2G\langle m \rangle)$ \citep{Merri13,merritt13}.
 After rescaling the simulations with Equations~(\ref{eq:rescale1}--\ref{eq:rescale2}), our system is characterized by an influence radius 
equal to the Milky Way's estimated value.
Assuming an  effective mass of $m_{\rm eff}\sim 1 \Ms$, the relaxation time at the Sgr A* influence radius is $\sim 20-30$ Gyr \citep{merritt2010}.
In direct $N$-body simulations, the relaxation time is reduced by a factor $m_{\rm sim}/m_*$ due to the smaller number of particles as compared to real stellar systems, where $m_{\rm sim}$ and $m_*$  are the simulated and actual single stellar masses, respectively. 
Therefore, in order to mimic the NSC's long-term evolution, we carried out our simulations up to a fraction $m_{\rm sim}/m_*$ of the observed relaxation time-scale. 
These approximations, widely used in the field of numerical simulations, alleviate the large $N$ problem 
($N\gtrsim 10^8$ in reality, see for instance \citealt{AMB,antonini13,perets14,tsatsi17}). 
Upon this approximation, we selected the snapshot corresponding to 12 Gyr to perform our analysis.

\subsubsection{Mass segregation}

The other important process to take into account is the possible imprint of mass segregation in the observational properties. Indeed, the MSP and CV progenitor stars may have already undergone mass segregation in their parent GC when they reach the Galactic Centre, possibly affecting their distribution within the NSC after formation.
The GCs' infall is regulated by dynamical friction, whose timescale depends on the mass of the GC as $\langle m\rangle/m_{\rm GC}$.
Similarly, mass segregation in dense clusters also operates in a fraction of the relaxation time-scale \citep{SPITZER,zwart02,AS16} 
\begin{equation}
t_{\rm seg} = \frac{0.65{\rm Gyr}}{{\rm ln}(0.4M/\langle m\rangle)}\left(\frac{M}{10^5\Ms}\right)^{1/2}\left(\frac{1\Ms}{\langle m\rangle}\right)\left(\frac{r_h}{1{\rm pc}}\right)^{3/2},
\label{segr}
\end{equation}
where $\langle m\rangle$ is the mean stellar mass and $r_{\rm h}$ is the cluster half-mass radius. 
Since $t_{\rm seg} \propto M^{0.5}$ while $t_{\rm df}\propto M^{-0.67}$ \citep{ASCD15He}, the lighter the cluster the higher the probability that it reaches the galactic centre in a mass-segregated state. This is an oversimplification of the problem, since the segregation process depends mainly on the internal properties of the cluster, e.g. core radius, metallicity, mass function, while the cluster infall depends on its orbital properties and the host galaxy structure. 
In our simulations we cannot account for this effect self-consistently, as we used single mass models for both the clusters and the background galaxy and our mass resolution is much larger than $1\Ms$.
As we will discuss in detail in Section \ref{dMSP}, in order to alleviate our blindness of the MSPs and CVs formation history, in our calculations we assume that these sources at formation are either segregated into the parent GC core or completely unsegregated. 
Since the mixing time in globular clusters is of order 10 times the relaxation time \citep{meiron18} and the outer stars are stripped more easily from GCs, the initial conditions of MSPs and CVs within GCs affects their final distribution in the Galactic bulge.

\subsection{Selecting MSP and CV candidates in $N$-body modelling}

Our $N$-body simulations are based on single-mass particle representations of both the infalling GCs and the galaxy nucleus, while stellar evolution and binary formation are not treated in any way. 
The number of particles used is sufficiently high to resolve the stellar distribution in the star clusters' cores and in the SMBH surroundings, thus providing a statistical basis sufficiently robust to measure the evolution of stellar orbits.
We select and label particles as MSP or CV candidates, as discussed in \ref{nMSP} and \ref{dMSP}, and follow their evolution from their initial position within the parent cluster up to their final position in the Galactic Centre after the NSC formation.
The selection procedure accounts for two important quantities:
(1) the fraction of stars that can form an MSP or CV in a massive GC; and (2) the level of mass segregation of the parent GC when it impacts the SMBH. 
We discuss these aspects in Sections \ref{nMSP} and \ref{dMSP}, respectively.

\section{Results and Discussion}
\label{results}

In this section we use our rescaled $N$-body results to investigate the role of the GCs' infall in the production of high-energy emission from the Galactic Centre. 
The bottom panel of Figure \ref{dens} compares the NSC surface density profiles for our three simulations. The most prominent difference between the models is that a central SMBH strongly affects the inner mass distribution:
with no SMBH (model S3), the central density is up to five times larger. On the other hand, the differences in the star clusters' initial conditions (models S1 and S2) have only a weak impact on the final matter distribution. Beyond $\sim$3 pc the three profiles become almost indistinguishable, due to the fact that above this length scale the dominant contribution comes from the background galaxy in this model.\footnote{Note that it may come from a large number of primordial GCs that sink to this region from a kpc distance, modelled as a Galactic contribution here \citep{brandt15}.} These findings are consistent with previous models of the Milky Way NSC by \cite{AMB}. In particular, our surface density profile matches their Figure 4 in both the central surface density and the NSC effective radius. 

The absence of significant differences between the density profiles on length scales larger than 10-100 pc could imply that the GC initial conditions do not play a significant role  as far as the gamma ray emission is concerned, which is observed with {\it Fermi} with a poorer resolution. 
The radial profile of hard X-ray emission with NuSTAR has a better resolution of the Galactic Centre, which might offer further constraints on the GC initial conditions.
The NuSTAR satellite has an angular resolution of $58''$ (HPD, corresponding to $2.2$ pc at 8 kpc), a FWHM equal to $18''$ (0.7 pc) and a field of view of $6'$ (14 pc).\footnote{\url{https://heasarc.gsfc.nasa.gov/docs/nustar/nustar_tech_desc.html}.}
In the next sections we investigate whether the level of mass segregation in the infalling clusters and the presence of the SMBH may alter the final distribution of both MSPs and CVs in the very inner region of the Galaxy.

\subsection{The expected number of MSPs and CVs in the infall scenario}
\label{nMSP}

In this section we provide a crude estimate of the number of MSPs and CVs expected to be found in the Galactic Centre under the hypothesis that the NSC formed by repeated GC infall.

\subsubsection{Number of MSPs in the NSC}

 We determine the number of MSPs based on the following phenomenological parameters. 
\begin{itemize}

\item $\mu_{\mathrm{GC}i}$: the initial fraction of mass in the $i^{\rm th}$ GC in the Galactic bulge compared to the total Galactic bulge mass. 

\item $\mu_{\mathrm{NSC,GC}}$: the final fraction of mass in the NSC comprised of GC debris.

\item $\mu_{\mathrm{NSC,G}}$: the fraction of mass in the NSC already present in the Galactic centre before NSC formation. We assume that $\mu_{\mathrm{NSC,GC}}+\mu_{\mathrm{NSC,G}}=1$. 

\item $\nu_{\mathrm{MSP}}$: the number of MSPs per unit mass in GCs, $\nu_{\mathrm{MSP}}=N_{\mathrm{MSP}}/M_{\mathrm{GC}}$.

\item $\delta_i$: the final fraction of the $i^{\rm th}$ GC mass in the bulge that makes it to the NSC. The rest $1-\delta_i$ represents the final fraction of GC mass deposited in the bulge due to GC evaporation and tidal disruption of GCs. Thus $\mu_{\mathrm{NSC,GC}}=\sum_i \delta_i \mu_{\rm GCi}$.
\item
$\eta_{\mathrm{MSP}}$: 
the ratio between the initial mass in MSPs in the Galactic field and that in GCs.
Since MSP formation is correlated with the dynamical encounter rate \citep{bahramian13,hui10}, the formation efficiency in the Galactic field is smaller than in GCs, $\eta_\msp<1$.
\end{itemize}
With these parameters, the total number of MSPs in the NSC can be expressed as
\begin{equation}
N_\msp = \nu_\msp \left(  \eta_\msp \mu_{\rm NSC, G} + \sum_i \delta_i \mu_{{\rm GC}i} \right)M_\nc,
\label{nmsp}
\end{equation}
where the sum is over all GCs in the Galactic bulge. We determine the parameters as follows.

\begin{table*}
\begin{center}
\caption{Number of expected MSPs in observed GCs.}
\label{GCMSP}
\begin{tabular}{lccccc}
\hline
GC name & $N_{\rm MSP0}$ & $M_{\rm GC}$ & $d$ & $F_{2{\rm ~Gev}}$ & ref.\\
 &&($10^6\Ms$)& kpc & ($10^{-9}\flux$) & \\
\hline
\hline
47 Tuc          &  33   &  0.700 & 4.5  &5.6  &\cite{marks2010}\\
$\Omega$ Cen    &  19   &  1.500 & 5.2  &2.8  &\cite{marks2010}\\
M 62            &  76   &  1.220 & 6.8  &3.8  &\cite{boyles2011}\\
NGC 6388        &  180  &  2.170 & 9.9  &3.4  &\cite{boyles2011}\\
Terzan 5        &  180  &  0.374 & 6.9  &12.6 &\cite{boyles2011}\\
NGC 6440        &  130  &  0.811 & 8.5  &2.9  &\cite{boyles2011}\\
M 28            &  43   &  0.551 & 5.5  &3.8  &\cite{boyles2011}\\
NGC 6541        &  47   &  0.572 & 7.5  &0.9  &\cite{boyles2011}\\
NGC 6752        &  11   &  0.140 & 4.0  &0.5  &\cite{marks2010}\\
M 15            &  56   &  0.560 & 10.4 &-    &\cite{marks2010}\\
\hline
\end{tabular}
\end{center}
\begin{tablenotes}
\item Col. 1 GC name. Expected number of MSPs \citep{abdo10}. Col. 3: GC mass. Col. 4: observed flux at 2 GeV \citep{cholis14}. Col. 5: GCs distances \citep{harris96}. Col. 6: reference for GC masses.
\end{tablenotes}
\end{table*}

We follow \citet{abdo10} to estimate $\nu_\msp$ using Fermi observations of the gamma-ray flux for ten old and massive GCs, which yields the enclosed number of MSPs, $N_{\rm MSP0}$.  Table \ref{GCMSP} summarizes $N_{\rm MSP0}$ and the mass of the host GCs, derived from literature. Our GC models have similar GC masses, and therefore assume similar numbers of MSPs. 
The \cite{abdo10} sample is comprised of old GCs, with ages $\sim 10$ Gyr. Hence, $N_{\rm MSP0}$ represents the lower limit of MSP progenitors. Neutron stars form over a time-scale $\sim 10-100$ Myr, with a substantial fraction of them being ejected due to supernovae kicks. On the contrary, MSPs form after a NS is captured in a binary system and the binary is hardened by dynamical encounters to be spun up into a MSP. This typical time is $\sim 1$ Gyr, larger than the time-scale for NS formation and ejection. Thus, we do not expect a significant variation in $N_{\rm MSP0}$.
We get 
$\nu_\msp = 775/(8.6\times 10^6\Ms) = 9 \times 10^{-5}$ MSP $\Ms^{-1}$.
We measure $\delta_i$, $\mu_{{\rm GC}i}$ and $\mu_{\rm NSC, G}$ from the simulation.

\begin{table*}
\caption{}
\centering{Main parameters for MSPs and IPs number calculation}
\begin{center}
\begin{tabular}{cccccccc}
\hline
\hline
source & $\nu ~(\Ms^{-1})$ & $f$ & $\delta$ & $\mu_{\rm GCS}$ & $\eta$ & $\mu_{\rm NSC, G}$& $N_{\rm src}$\\ 
\hline
MSPs & $9 \times 10^{-5}$  & $-$   & $0.5$& $1.3$& $0.0$& $0.5$& $1911$\\
IPs  & $1.92\times 10^{-3}$& $0.08$& $0.5$& $1.3$& $0.0$& $0.5$& $3255$\\
\hline
\end{tabular}
\end{center}
\label{tabn}
\end{table*}

The main parameter values used in the above calculations are summarized in Table \ref{tabn} (see top row for CVs). Assuming $\eta_\msp=0$ 
and substituting the parameters in \citet{abdo10} and Table~\ref{GCMSP} into Equation~(\ref{nmsp}),  we obtain $N_\msp = 1000$--$1200$ within 10 pc from Sgr A*, a number in good agreement with semi-analytic calculations and numerical modelling presented in the literature \citep{bednarek13,brandt15,abbate17}. We run calculations with $\eta_{\rm MSP}=0$, 0.01, and 0.1.
Here $\eta_{\rm MSP}\sim 0.01$ is compatible with observational evidence of a much smaller number of MSPs per unit mass in the Galactic field, up to $\sim 100$ times that in GCs \citep{grindlay09}.

\subsubsection{Number of CVs in the NSC}
We use a simple phenomenological model to derive the number of IPs in the Milky Way's NSC, which may be responsible for the hard X-ray emission observed in the Galactic Centre:
\begin{equation}
N_{\rm IP} = \nu_\cv f_{\rm IP}  \left(  \eta_\cv \mu_{\rm NSC, G} + \sum_i \delta_i \mu_{\rm GCi}\right)M_\nc.
\label{ncv}
\end{equation}
Here, $f_{\rm IP}$ denotes the fraction of IPs among all CVs. Based on observations of the ROSAT Bright Survey, \cite{pretorius13} estimated that a fraction $f_{\rm mCV}=0.2$ of all CVs in the Solar neighbourhood are magnetic, and about 40\% of the local magnetic CVs are IPs. Thus, we assume that $f_{\rm IP} = 0.4 f_{\rm mCV} = 0.08$. 

Further \cite{ivanova06} used Monte-Carlo models of star clusters and found the formation of $N_{\rm CVo} = 2490$ CVs in 13 massive GCs with lifetimes larger than $12$ Gyr. The total mass of GCs was $M_{\rm GCo} = 1.3\times 10^6\Ms$. Thus, we get, $\nu_\cv = N_{\rm CVo}/M_{\rm GCo} = 1.92\times 10^{-3}$ CV $\Ms^{-1}$. 

We use the same $\delta_i$, $\mu_{{\rm GC}i}$ and $\mu_{\rm NSC, G}$ measured from the simulation as for the MSPs.
The parameter values in Equation (\ref{ncv}) are summarized in Table \ref{tabn} (see bottom row for IPs). 
According to Equation \eqref{ncv}, the total number of IPs formed in the infalling clusters is $N_{\rm IP,tot} = 5994$. After the NSC build-up, the expected number of IPs, solely coming from the infalling clusters and deposited inside the inner 10 pc, is $N_{\rm IP} = 1823$. Given that the Galaxy's NSC effective radius is $r_{\rm NSC}\sim 4.2$ pc \citep{schodel14} and its radial inner slope is $\gamma_{\rm NSC} \simeq 1-2$ \citep{schodel14}, we can calculate the IPs' mean density assuming that the NSC distribution follows a powerlaw $n_{\rm IP}\propto r^{-\gamma}$. This leads to  $n_{\rm IP} = 0.35 - 1.6$ pc$^{-3}$. 
In the next section, we will show that this rough estimate agrees with the IPs' simulated radial distribution (Figure~\ref{dCV}), and it is well below the upper bound on the IP density inferred from observations \citep[$n_{\rm obs} \simeq 1-3$ pc$^{-3}$,][]{heard13,perez15,hailey16}. Thus, the simulated number of IPs 
within 150 pc is consistent with the observationally inferred values within the theoretical uncertainties. Further, we show in the next section, that the surface density and X-ray flux found in our simulations are consistent with observations, suggesting that a dry merger origin of the Galactic NSC is viable to explain the origin of the large population of MSPs and CVs that generate the observed gamma and X-ray emission in the Galactic centre.

\subsection{Mock catalog of gamma and X-ray sources in $N$-body simulations}
\label{dMSP}

In the following we combine the calculation performed to obtain the expected number of MSPs and CVs in our modelled clusters with the data provided by the numerical simulations to study the shape and characteristics of the $\gamma$ and X-ray emission expected from the Galactic centre.

Our $N$-body models have a sufficiently large number of particles to ensure a reliable selection of source candidates.
In the following, we will focus on model S2, due to the absence of big differences between S1 and S2 surface density profiles which represent spherical and planar initial GC distributions respectively (see Figure \ref{dens}). 
We will use models S1 and S3 in the next sections to highlight the role played by the central SMBH. 

In each cluster, we randomly selected $\nu_j M_{{\rm GC}j} $ particles, where the subscript $j$ refers either to MSPs or CVs.  
The number of sources for each cluster, rescaled to the Milky Way nucleus, is obtained through Equations (\ref{nmsp}) and (\ref{ncv}). 
 All particles in our simulation have the same mass, so any treatment of mass segregation of the MSPs and CVs can only be done in postprocessing.  We qualitatively account for this effect by preferentially selecting tracer particles within or beyond a given radius from the cluster center.  We first fix a radius $l$ in units of the core radius, and then vary the fraction of MSPs and CVs tracing the mass inside ($f_s$) and outside ($1-f_s$) of $l$.  For example, $f_s=0.5$ and $l=1$ corresponds to half of the MSPs and CVs tracing mass in the core and the other half outside the core, while $f_s=1$ and $l=\infty$ corresponds to all MSPs and CVs tracing the cluster's stellar mass.  With $f_s=1$ and $l=1$, all of the MSPs and CVs are assumed to have mass-segregated into the cluster's core.

We do not account for strong encounters, as our numerical code does not implement any treatment for stellar binaries or tight multiple systems.  Table \ref{MOD} shows our assumptions on the MSP and CV populations. 
 Our approach assumes that the MSP and CV populations do not mix with the bulk of the cluster population after they have mass-segregated.  \citet{meiron18} showed that the mixing time of objects in the cluster is about 10 relaxation times, which is typically larger than the timescale on which the clusters are stripped by galactic tides. 

We generated five different models, characterised by different levels of segregation for MSPs and CVs. In these models we also varied the efficiency factors $\eta_{\rm MSP}$ and $\eta_{\rm CV}$, defined in Eqs. (\ref{nmsp}) and (\ref{ncv}), in order to outline the role of sources born in  the Galactic disk. For instance, models MSPa/CVa and MSPb/CVb refer to a population of MSPs and CVs confined initially to the cores of their parent clusters. The contribution formed in the Galactic centre is set to $\eta_{\rm MSP}=\eta_{\rm CV}=0$ in MSPa/CVb and $1\%$ in MSPb/CVb. In models MSPc/CVc, MSPd/CVd, MSPe/CVe we set the fraction of sources in the core to be $f_{\rm s}=50\%$. Model MSPf/CVf represents an unsegregated populations of MSPs and CVs.

For reference, models CVa refers to a completely segregated population of CVs in their parent clusters, while the contribution coming from CVs formed in the Galactic bulge outside of GCs is set to zero.
Moreover, Table \ref{MOD} summarizes our choices fot the Galactic field contribution, 
number of sources and fraction of sources contained within a given fraction of the cluster core radius.

\begin{table}
\centering
\caption{MSP and CV distributions in our simulations}
\begin{center}
\begin{tabular}{ccccc}
\hline
  & $f_s$ & $l$   &    $\eta$  & $N_{\rm src}$  \\
  &       & $R_c$ &            &     \\
\hline
 MSPa	& 1   & 1      & 0     & 2254   \\
 MSPb	& 1   & 1      & 0.01  & 2758   \\
 MSPc	& 0.5 & 1      & 0     & 2254   \\
 MSPd	& 0.5 & 1      & 0.1   & 6953   \\
 MSPe	& 0.5 & 0.2    & 0     & 2254   \\
 MSPf	& 1   & $10^3$ & 0     & 2254   \\ 
\hline 
 CVa    & 1   & 1       & 0     & 5994  \\
 CVb    & 1   & 1       & 0.01  & 59577 \\ 
 CVc    & 1   & $10^3$  & 0     & 5994  \\
 CVd    & 1   & $10^3$  & 0.01  & 59577  \\
 CVe    & 1   & $10^3$  & 0.002 & 16710  \\
\hline
\end{tabular}
\end{center}
\begin{tablenotes}
\item Col 1. Model name. Col. 2: fraction of sources within $l$ times the cluster core radius, $R_c$. Col. 3: radius, in unit of the cluster core radius, within a fraction $f_s$ of the sources is enclosed; $10^3$ means that sources are distributed wherever inside the cluster tidal radius. Col. 4: efficiency factor as defined in Eqs. \ref{nmsp} and \ref{ncv}. Col. 5: total number of sources. 
For models with $\eta>0$ (including sources formed in the Galactic field), 
the numbers show sources within the modelled region of the bulge, $r<150$ pc.
\end{tablenotes}
\label{MOD}
\end{table}

\subsection{Comparison of observations with simulations}

The power emitted in the 2 GeV band from a single MSP can be estimated as
\begin{equation}
L_{\rm 2 GeV} = \frac{F_{\rm 2 GeV}(4\pi D^2)}{M_{\rm GC}} m_{\rm MSP},
\end{equation}
where $m_{\rm MSP}$ is the MSP mass, $F_{\rm 2 GeV}$ the GC observed flux, $M_{\rm GC}$ its mass and $D$ its distance from the observer. Following \cite{abdo10}, we found $L_{\rm 2 GeV} \sim 4\times 10^{35}\flux$ per MSP.

\begin{figure}
\centering
\includegraphics[width=8cm]{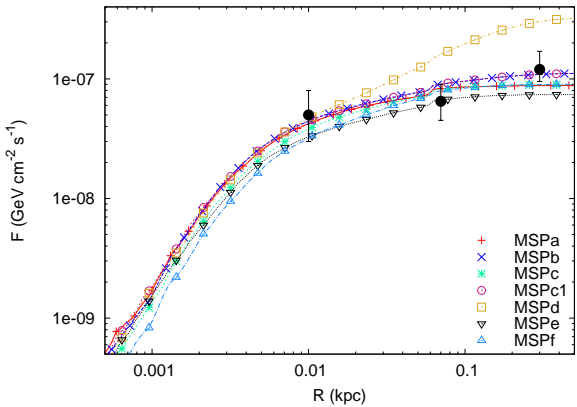}
\caption{Flux for different MSP models for the  initially disk-like distribution of globular clusters (model S2) and different assumptions on the initial distribution of MSPs within the cluster and the number of MSPs in the galactic field. For the definition of models see Table~\ref{MOD}. The large deviation for model MSPd is due to the assumed high Galactic field MSP contribution $N\sim 4000$ MSPs ($\eta=0.1$). Note that the modelled region is meaningful within 150 pc (the sampling is exponentially truncated beyond 150 pc). The MSP population from infalling GCs is also underestimated further out due to the neglect of infalling GCs from outside of 200 pc. The black filled dots represent observed $\gamma$-excess reported by \citet{brandt15}.
}
\label{MSPcom}
\end{figure}

\begin{figure}
\centering
\includegraphics[width=8cm]{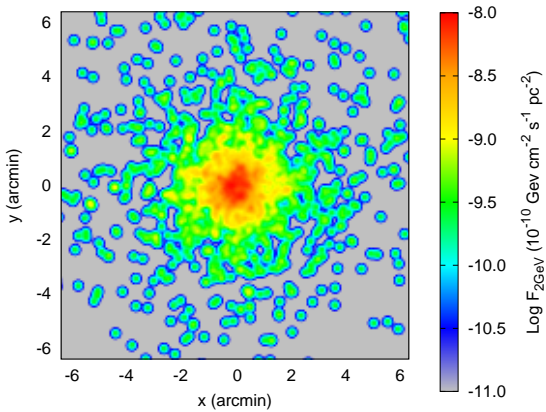}\\
\caption{Surface map of the 2 GeV intensity produced by the Milky Way's NSC as calculated from our simulation S2, model MSPa.
}
\label{map2gev}
\end{figure}

Figure \ref{MSPcom} shows the observed flux in all the configurations tested as a function of radius and Figure \ref{map2gev} shows the 2D surface map of the gamma-ray intensity emitted by the Galaxy's NSC in all the models investigated.  Different lines in Figure \ref{MSPcom} show different assumptions on the initial internal distribution of MSPs within the globular clusters and the initial number of MSPs outside of globular clusters in the Galaxy as shown by Table~\ref{MOD}. We find that many of these models are in tension or inconsistent with observations. In particular, if the initial fraction of MSPs in the Galaxy is high $\eta_{\rm MSP}=0.1$ (Model MSPd), the simulation greatly overpredicts the gamma-ray flux at 80 and 200 pc relative to the observed values. Further, an initially unsegregated population of MSPs in model MSPf produces a low flux in the inner region of the Galaxy $<1$ pc at the margin of the observational error. 
If all the MSPs are contained within their host clusters' core and the galaxy does not contribute to their population at all (model MSPa), the expected flux is consistent with observations between 10 and $\sim 80$ pc, while it is $7\%$ smaller than the flux observed at 150 pc.
However, we must stress here that our numerical model for the galactic bulge extends up to 150 pc at most, being exponentially truncated outward. Thus, the GC material delivered to 150pc from the outer regions is underestimated in the simulation.


The best agreement between observations and the simulations shown in Figure \ref{MSPcom} is achieved by model MSPb, which is characterised by having all the MSPs contained within the core radius of their host clusters and having a very small contribution of sources formed in the Galactic nucleus, namely $\eta_{\rm MSP} = 0.01$.
Note that in this model $1912$ MSPs were brought to the Galactic Centre by GC infall, while the progenitors of $503$ MSPs were born in the Galactic Centre before the GC merging process occurred.
Thus, our results suggest that before NSC formation the MSPs reside within the core radius of their host clusters and only $\sim 25\%$ of the total number of MSPs currently in the Galaxy's NSC have formed in the Galactic nucleus.
However, note that several other models may also be statistically consistent due to the current level of large observational errors shown in Figure \ref{MSPcom} and due to the limitations of our models neglecting GCs initially further out. 

For instance, the observed gamma-flux is well-fit by model MSPc, which assumes partial mass-segregation within their parent clusters. However, in the extreme limit in which MSPs are completely unsegregated (model MSPf), the resulting flux is much lower than observations.
Our results suggest that MSPs were at least partly segregated when they reached the inner galactic regions.

This has interesting implications on the dynamics of the parent clusters. Indeed, 
according to Equation \ref{segr}, stars having a mass in the range $\sim 10\Ms$ will segregate into the parent cluster centre in $t_{\rm seg} \simeq 15$ Myr, assuming a cluster mass of $10^6\Ms$. Given the similarity between the $t_{\rm seg}$ and the NSC assembly time-scale, $t_{\rm NSC}\gtrsim 100$ Myr, we expect that the population of MSPs progenitors arrived at the Galactic Centre at least partially segregated, although $t_{\rm seg}$ provides only a crude estimate of the actual segregation time-scale.

\begin{figure}
\centering
\includegraphics[width=8cm]{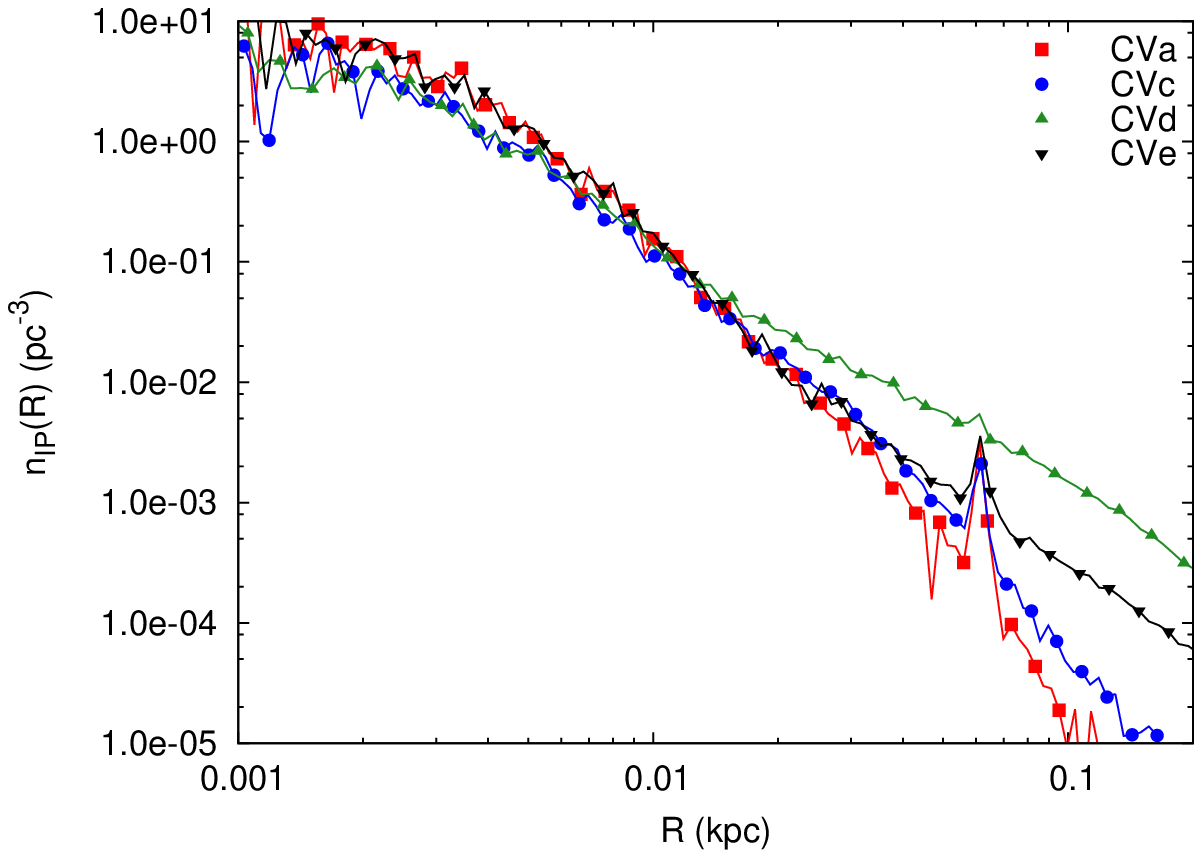}\\
\includegraphics[width=8cm]{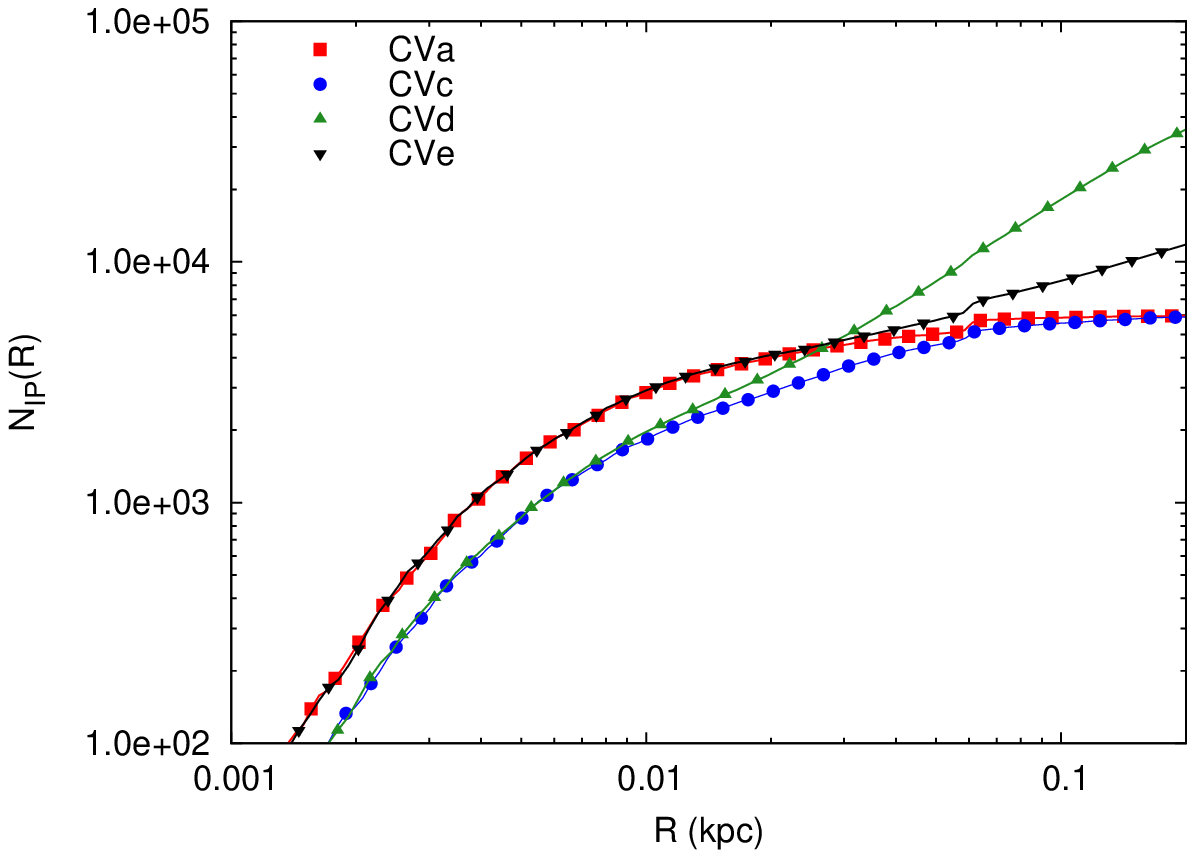}\\
\caption{Top panel: number density profile of CVs in different configurations. Bottom panel: CV cumulative distribution profile. 
}
\label{dCV}
\end{figure}

Similarly to the study of MSPs, we investigate the role of mass segregation and varying fraction of Galactic CVs using models listed in Table \ref{MOD}.
Figure \ref{dCV} shows the number density distribution of CVs and their averaged density per total volume $\langle n_{\rm CV}\rangle$. 
The difference between the IPs' distribution indicates how the initial IP distribution affects their final density profile after NSC formation.
As expected, an initially unsegregated IP population is characterised by a less centrally concentrated distribution of IP population in the NSC. 
Note in Figure \ref{dCV} that model CVe is characterised by a cored distribution inside $2$ pc, while model CVa has a steeper distribution $n(r)\propto r^{-\gamma}$, with slope $\gamma \sim 0.32\pm0.03$.
Comparing models CVa and CVc initially without Galactic IPs with models CVd and CVe for which $\eta_{\rm CV} = 0.01$ (see Table \ref{MOD}) shows that in the latter case the CV population formed in the Galactic centre dominates outside of $\sim 12$ pc. However, within this radius GCs may deliver a dominant population of IPs. 
\cite{hailey16} derived an IP number density of $\langle n_{\rm IP}\rangle \sim 2-10$ pc$^{-3}$ in the inner 10 pc to match NuSTAR observations, while our simulation models predict an IP density $\langle n_{\rm IP}\rangle \sim 0.2$ pc$^{-3}$, an order of magnitude smaller than the number observations suggest. Nevertheless, this discrepancy does not necessarily rule out the IP interpretation of the NuSTAR data. Indeed, we argue that, the X-ray surface density profile inferred from our simulations is compatible with observations, as well as the simulated X-ray flux morphology is quite similar to the observed map.
The discrepancy might be due to the strategy followed in \cite{hailey16} to infer the number of IPs from the observed number of main sequence stars in the Galactic Centre. In particular, they assume that all the binaries containing a white dwarf are CV (see their Sect. 7.2), which amounts to an upper limit, since only tight binaries can lead to the formation of a CV. Assuming that the IP interpretation of the hard X-ray foreground is correct, our calculations suggest that at most $10\%$ of these binaries undergo a CV phase.
We note that \cite{pretorius13} suggested an IP density $n_{\rm IP} = 6.4\times 10^{-4}$ pc$^{-3}$ calculated within a sphere with radius 150 pc. In our models the number of IPs enclosed within this radius yields to an average density $n_{\rm IP} = 1.4-6.7\times 10^{-4}$ pc$^{-3}$, depending on the choice of $\eta_{\rm IP}=0,~0.01,~0.002$.

\begin{figure}
\centering
\includegraphics[width=8cm]{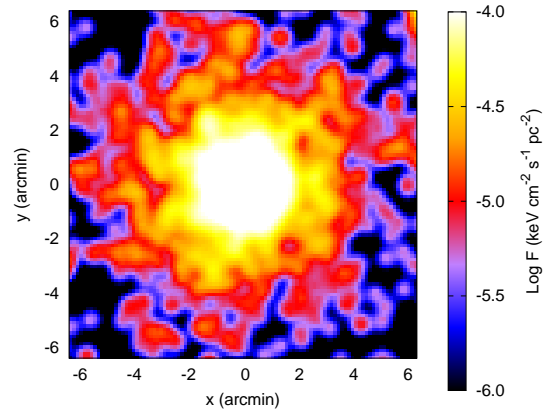}\\
\caption{
Emitted flux from IPs in our model. 
}
\label{CVmap}
\end{figure}

We show in Figure \ref{CVmap} the surface flux map in our CVb model. We limited the field of view in this case to 6.4 arcmin, in order to compare with observations provided by NuSTAR \citep{mori15}. 
A qualitative comparison with the inner galactic regions in the 20-40 keV energy range (Fig. 5 \cite{mori15}, shows similarity between the X-ray image from our simulations, rescaled to the MW centre,  and the observed one. Note that the simulated IPs' morphology is qualitatively consistent with the observed X-ray image, although a more rigorous comparison is difficult due to the fact that the observed features are sensitive to the initial conditions that affect NSC formation. Interestingly spiral streamlike structures are visible in the simulation image (see e.g. between $-5'<x<-3'$ and $-2'<y<0'$).

\begin{figure}
\centering
\includegraphics[width=8cm]{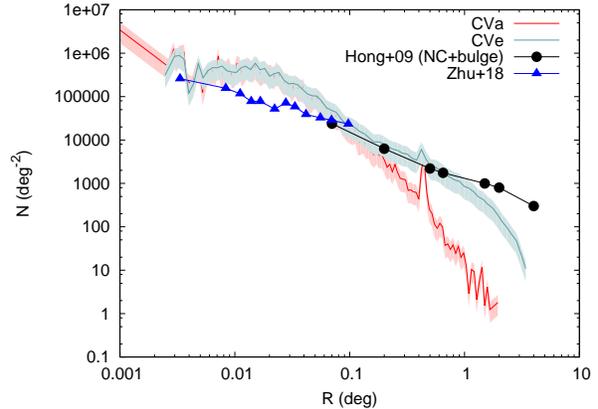}
\caption{Surface X-ray profile calculated for model CVa (red line) and CVe (blue line). The filled black circles represent observational data by \citet{hong09} (their Fig. 6), where we included only the contribution from the NSC and the Galactic bulge. The filled grey triangle are data adapted from \citet{zhu18}, based on Chandra observations of X-ray sources in the Galactic Center. Note that 1 deg corresponds to 140 pc.}
\label{Xflux}
\end{figure}

Figure \ref{Xflux} shows the number of sources per square degree in our model, compared to observational results from \cite{hong09} (their Figure 6). Since our models take into account only the NSC and the Galactic bulge, we show only these two contributions from \cite{hong09}.
We found an overall agreement between 0.07 and 1 deg, corresponding to $\sim 10-140$ pc at 8 kpc. The discrepancy outside of 1 deg is due to the adopted exponential truncation in our model, which makes the simulation unreliable outside the truncation radius (150 pc).
In particular, the best agreement with observations is achieved with model CVe, in which we assumed $\eta = 2\times 10^{-3}$. In this model, 2/3 of the total number of IPs within the inner 150 pc formed in the Galactic field, while the remaining population originated in star clusters with radially segregated initial profiles.
We also compared our models with Chandra data \citep{zhu18}, finding a discrepancy in the emission from within 0.1 deg. This difference can be due to our assumption that IPs are the only sources emitting in the X-ray band and that all the IPs in our sample are characterised by a luminosity of $10^{32}$ erg s$^{-1}$. On the other hand, the discrepancy might imply that something in our knowledge of CVs and IPs formation processes is missing. Indeed, it is possible either that i) the number of CVs in GCs is smaller than inferred from observations, or ii) that the fraction of CVs that turn into IPs is smaller than expected.
 For instance, a recent paper based on Chandra observations of Galactic GCs suggests that the population of CVs forming in GCs could be small compared to the field \citep{cheng18}.  

\begin{table}
\centering
\caption{Number of CVs over total number of stars}
\begin{center}
\begin{tabular}{ccc}
\hline
 & $\Gamma_{10}$ & $\Gamma_{100}$ \\
 & $r<10$ pc & $r<100$ pc \\
\hline 
 CVa    &  $1.69\times 10^{-4}$  &  $2.69\times 10^{-5}$ \\
 CVb    &  $1.79\times 10^{-4}$  &  $8.10\times 10^{-5}$ \\ 
 CVc    &  $1.10\times 10^{-4}$  &  $2.57\times 10^{-5}$ \\
 CVd    &  $1.14\times 10^{-4}$  &  $8.03\times 10^{-5}$ \\
 CVe    &  $1.09\times 10^{-4}$  &  $3.65\times 10^{-5}$ \\
\hline
\end{tabular}
\end{center}
\begin{tablenotes}
\item Col 1. Model name. Col. 2: fraction of CVs over the fraction of stars in a 10 pc volume. Col. 3: same as column 2, but in a 100 pc volume.
\end{tablenotes}
\label{fff}
\end{table}

\citet{hong09} calculated the number of CVs within $\sim 1$ kpc, normalized to the total number of stars, needed to ascribe the Milky Way's hard X-ray emission to IPs. They found that this quantity should be in the range $\Gamma_{\rm IP}\sim (1.6-9.5) \times 10^{-5}$. In order to compare with observations, we calculate the same fraction, assuming that IPs are only $0.8\%$ of the whole CV population \citep{pretorius13}. We find that in the inner 100 pc, $\Gamma_{100}=(2.5-8)\times 10^{-5}$, in agreement with the predictions based on observations. More recently, \cite{zhu18} reported an enhanced abundance of CVs in the central 10 pc with respect to the outside by a factor 2, compatible with our findings summarized in Table \ref{fff}.
Thus, we conclude that the population of IPs dominating the X-ray emission in the Galactic centre could have mostly originated in GCs.

\section{Discussion}
\subsection{The link between the NSC formation history and the Galactic centre BH population}

Our approach shows that many compact sources can be deposited into the Galactic Centre in the course of the NSC's formation.
The recent discovery of 12 low-mass X-ray binaries (LMXBs) orbiting around 1 pc from Sgr A* \citep{hailey18}, raised further questions about the evolution of our Milky Way centre. As discussed by \cite{generozov18}, these sources might have formed in tidal capture by single BHs in the dense environment characterizing the NSC.

A ``wet'' NSC origin, in which the stars formed in situ, will leave a large population of BH remnants near the Galactic centre.  In this section we explore whether a dry-merger scenario can produce a population of remnants compatible with the inferred BH population, which could number as high as 20,000 \citep{MiraldaEscude00}.

 We use the same approach as for MSPs and CVs to infer the number of BHs delivered to the NSC by infalling clusters:
\begin{equation}
N_\bh =\Gamma_{\rm ret}\nu_\bh\left(\eta_\bh\mu_{\rm NSC,G} + \sum_i \delta_i \mu_{\rm GC,i}\right)M_\nc
\end{equation}
Here, $\Gamma_{\rm ret}$ is the BH retention fraction.  BHs formed in clusters may be ejected immediately due to a large natal kick, or they may be ejected later due to dynamical interactions as they mass segregate to the core and undergo strong scatterings. The retention fraction here is the fraction of BHs that remain bound to the cluster until it reaches the NSC.  Although uncertain, this parameter is thought to be $\gtrsim$0.5 on both numerical \citep{repetto15,morscher15,mandel16,peuten16} and observational \citep{strader12,chomiuk13,miller15,bahramian17,giesers18} grounds.  We assume a \cite{kroupa01} initial mass function and calculate $\nu_\bh$ as the number of stars with initial masses above $18 \Ms$.  In this way we obtain a ratio of BHs to stellar mass of $\nu_\bh \simeq 3.5\times 10^{-3} \Ms^{-1}$.

We measure the fraction of GC mass transported to the galactic centre $\delta_i$ directly from the simulations, while we assume that the efficiency of BH formation is the same both in the galaxy field and the cluster, thus implying $\eta_\bh = 1$.

Assuming a $\Gamma_{\rm ret} = 0.5$ retention probability, our results suggest that GCs deposit 
\begin{equation}
N_\bh = 2.4\times 10^4 \nonumber
\end{equation} 
BHs into the NSC, while a similar number of BHs should form directly in the galactic nucleus while the NSC grew up. 
Once deposited into the Galactic Centre, these BHs will segregate into the deepest NSC regions due to dynamical friction, which operate very efficiently for heavy objects like stellar mass BHs, becoming the most likely progenitors for the observed population of LMXBs.

\subsection{Implication of a NSC wet origin for the $\gamma$ ray excess}

If the NSC formed according to the wet scenario, its formation would have occurred by gas fragmentation around the SMBH. 
In this case, the NSC would behave similarly to a very massive and dense star cluster.
 Assuming a total mass $M_\nc = 2.5\times 10^7\Ms$ and half-mass radius $r_h\simeq 2$ pc, the NSC is expected to undergo mass segregation in a fraction of its relaxation time, $\sim 200$ Myr according to Equation \eqref{segr}. This is a clear oversimplification since the NSC will grow in time, thus implying that $M_\nc$ and $r_h$ are time-varying quantities. Therefore, we caution that these are rough estimates, and represent a useful point of comparison for the next generation of numerical models.  These models will hopefully have sufficient resolution to discern the motion of actual MSPs or CVs candidates in galactic nuclei.

The expected population of MSPs in a NSC formed entirely in-situ is uncertain. Escape speeds from the NSC are much higher than for GCs, so the NSC should retain a larger fraction of its neutron stars. Higher velocity dispersions also reduce the rate of strong encounters and may therefore inhibit MSP formation.  It is not clear whether these effects combine to increase or decrease the MSP population per unit stellar mass relative to that seen in GCs \citep{FaucherGiguere11,dexter14}. As a simple baseline model, we assume that they cancel out and produce a similar MSP abundance per unit mass as that seen in GCs, or $N_{\rm MW} \sim 2450$ MSPs.  We then produce gamma ray predictions from the wet formation scenario by randomly selecting $N_{\rm MSP,wet} \equiv  N_{\rm MW}$ particles within the spatial region enclosing the NSC.

This assumption for the wet formation scenario directly implies that the MSPs' radial distribution follows the NSC radial distribution. 
Again, using these simulations, we cannot directly account for mass segregation of stars in the NSC, so we assume that a fraction $n_{\rm MSP,wet}$ of stars is enclosed within $R_{\rm max}$ times the NSC core radius $r_{\rm cNSC}$, which for the Milky Way's NSC is $r_{\rm cNSC} \simeq 1$ pc.

We investigated two different cases: i) the whole population of MSPs is fully segregated inside the NSC core radius, i.e. $R_{\rm max}=r_{\rm cNSC}$ and $n_{\rm wet} = 1$; ii) the MSPs are distributed within $R_{\rm max} = 10r_{\rm cNSC}$ and $n_{\rm wet} = 1$. 

\begin{figure}
\centering
\includegraphics[width=8cm]{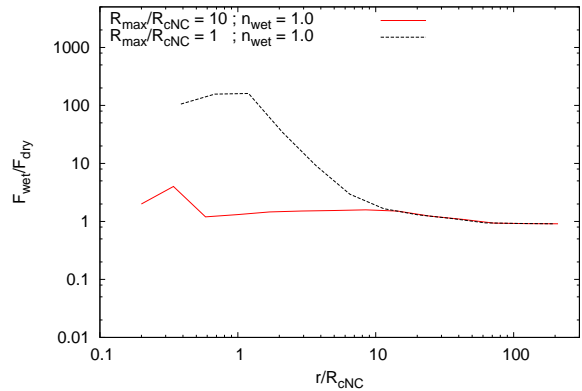}
\caption{Ratio between the MSPs $\gamma$ fluxes in the wet and dry NSC formation scenarios, for two different levels of MSPs segregation in the NSC formed in-situ.}
\label{wet}
\end{figure}

Figure \ref{wet} shows the gamma ray flux calculated in the in-situ scenario, $F_{\rm wet}$, normalized to the values obtained for model MSPa under the dry-merger scenario assumptions, in the three cases investigated.

Surprisingly, we found significant differences between dry- and wet- $\gamma$ fluxes emitted from the inner 10 pc. If the MSPs population is completely segregated in the ``in-situ NSC'', we found that the flux is up to 100 times those emitted from MSPs delivered from orbitally segregated star clusters.
However, if we assume that the whole population of MSPs is mixed within the inner 10 pc and follow the same radial distribution of background stars, the dry- and wet- scenarios produce similar results.

Since the mass of MSP progenitors is significantly higher than the average stellar mass in stellar system, they are expected to segregate into the NSC's central region over a dynamical friction time-scale (df), provided that this is shorter than the stellar lifetime. Indeed, the zero age main sequence mass range of a star which evolves into a neutron star is $7-20\Ms$ \citep{belczynski08}, and the average stellar mass is $m_{\rm eff}\sim 0.6\Ms$ for a \cite{kroupa01} mass function. The stellar lifetimes may be calculated using the SSE code for modelling stellar evolution \citep{hurley00},
which gives 56 Myr for $7 \Ms$ and 11 Myr for $20 \Ms$. 
Following \cite{AS16} (but see also \cite{ASCD14} for details), and assuming that the MSP progenitor stars were orbiting within the NSC core radius $r_* \simeq r_\nc$, on a nearly circular orbit, the df time can be calculated as:
\begin{equation}
t_{\rm df}(m_*) \simeq 0.3 {\rm Myr} \,g(e,\gamma_\nc)\left(\frac{m_*}{M_\nc}\right)^{-0.67},
\end{equation}
where $g(e,\gamma_\nc)$ is a smooth function of the NSC inner density slope and the star orbital eccentricity as given by \citet{ASCD15He}. Substituting the properties of our NSC, we find $t_{\rm df}(7\Ms) = 31$ Myr and $t_{\rm df}(20\Ms) = 15$ Myr, thus comparable to the stellar evolution time-scale above. This suggests that it MSPs progenitors can be partially segregated into the NSC, if formed in-situ therein.

\subsection{The role of a central SMBH}

In this section we focus on gamma-ray emission, to determine the role played by the GCs' initial conditions with and without an SMBH (S1, S2, S3, see Section \ref{model}).

\begin{figure}
\centering 
\includegraphics[width=8cm]{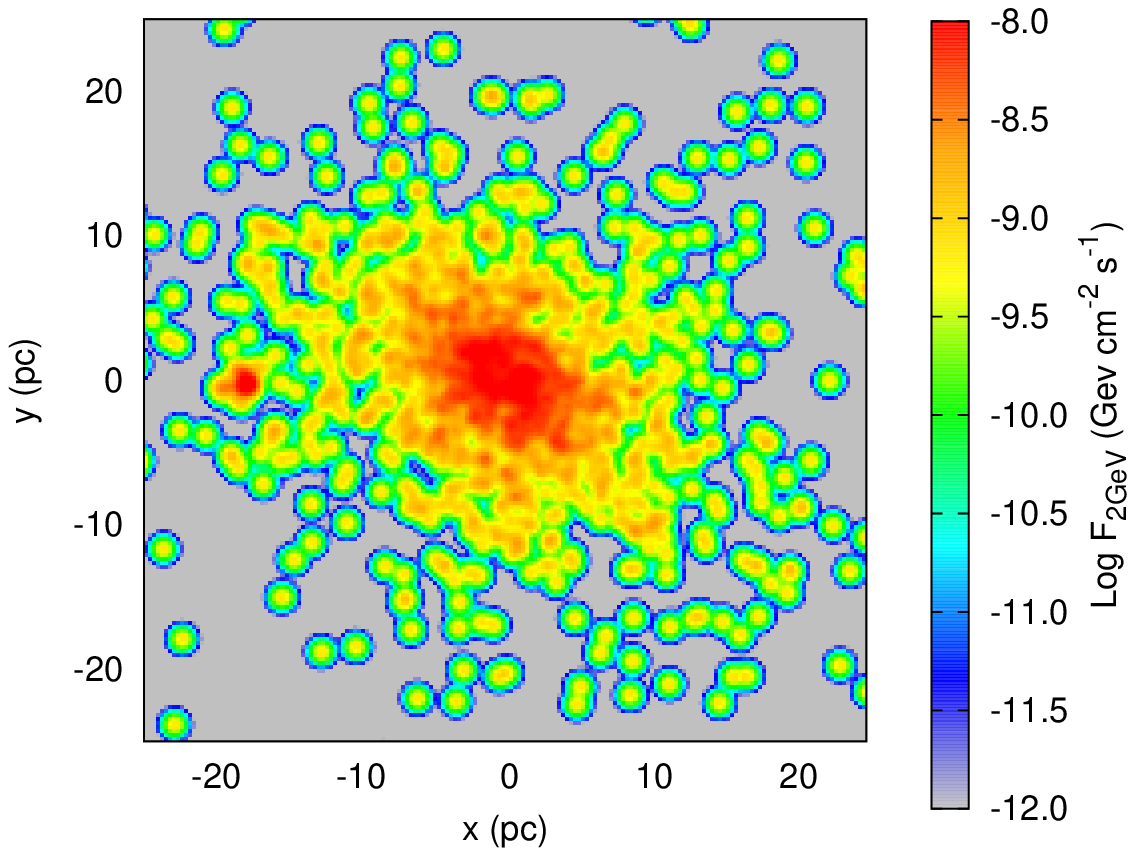}\\
\includegraphics[width=8cm]{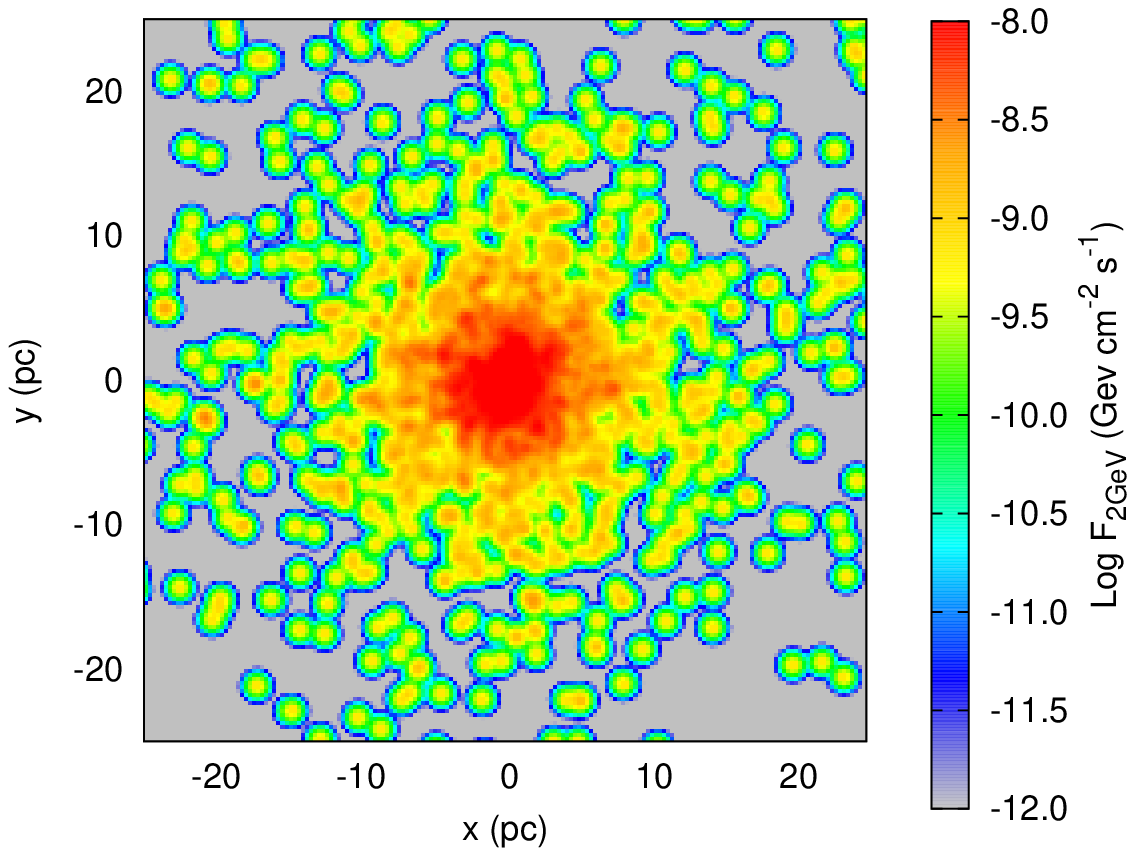}\\
\includegraphics[width=8cm]{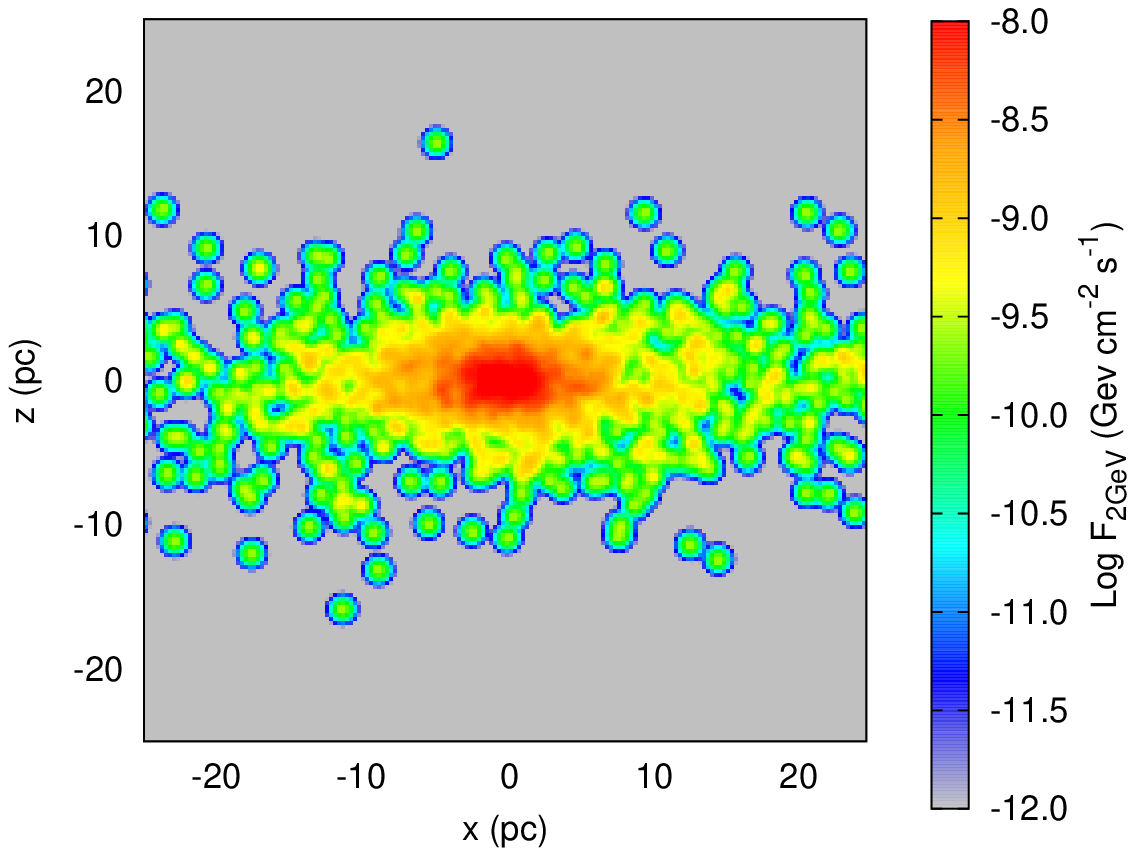}\\
\includegraphics[width=8cm]{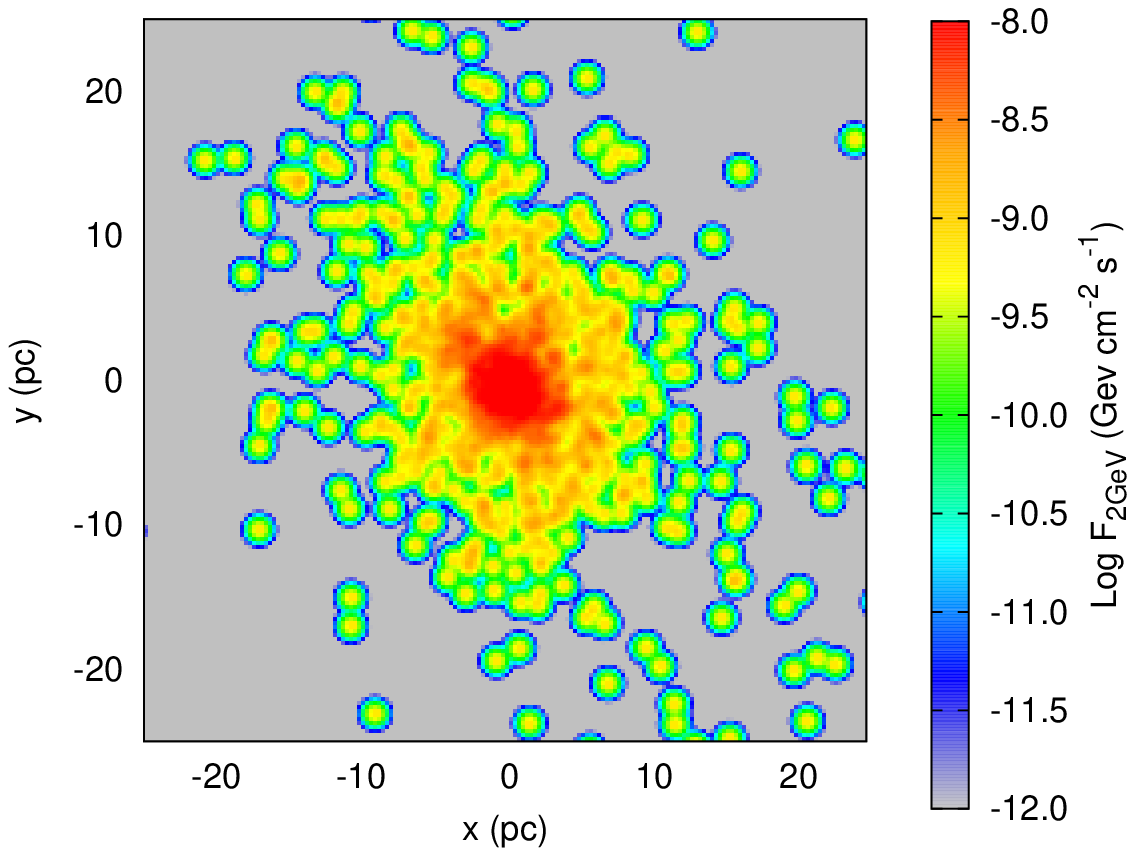}
\caption{Surface flux map in the three models investigated. The panels show respectively model S1, model S2 in the xy plane, model S2 in the xz plane, and model S3.}
\label{profiles}
\end{figure}

Figure \ref{profiles} shows the Galactic centre emission profile. The map shows the inner $25\times25$ pc region or 11\,arcmin$\times 11$\,arcmin around Sgr A*.
The Galactic centre MSP flux morphology, shows differences among different star cluster initial distributions.

The NSC flattening ratio $q_f$, calculated as the ratio between the minor and major axis of the ellipsoid enclosing the NSC, varies depending on the model considered. In our model S1, where clusters initial orbits have different orientations, we found $q_f = 0.6-0.8$, independently of the plane considered. It is worth noting that this value is really close to the Galactic NSC observed flattening \citep{schodel14,Chatzopoulos15,fritz16}. In the case of model S2, instead, the flattening ratio is smaller, $q_f\sim 0.52$, if we look in the plane perpendicular to the GCs initial orbital plane, while it rises up to $q_f=0.7-0.9$ if we look in the parallel plane.

\begin{figure}
\includegraphics[width=8cm]{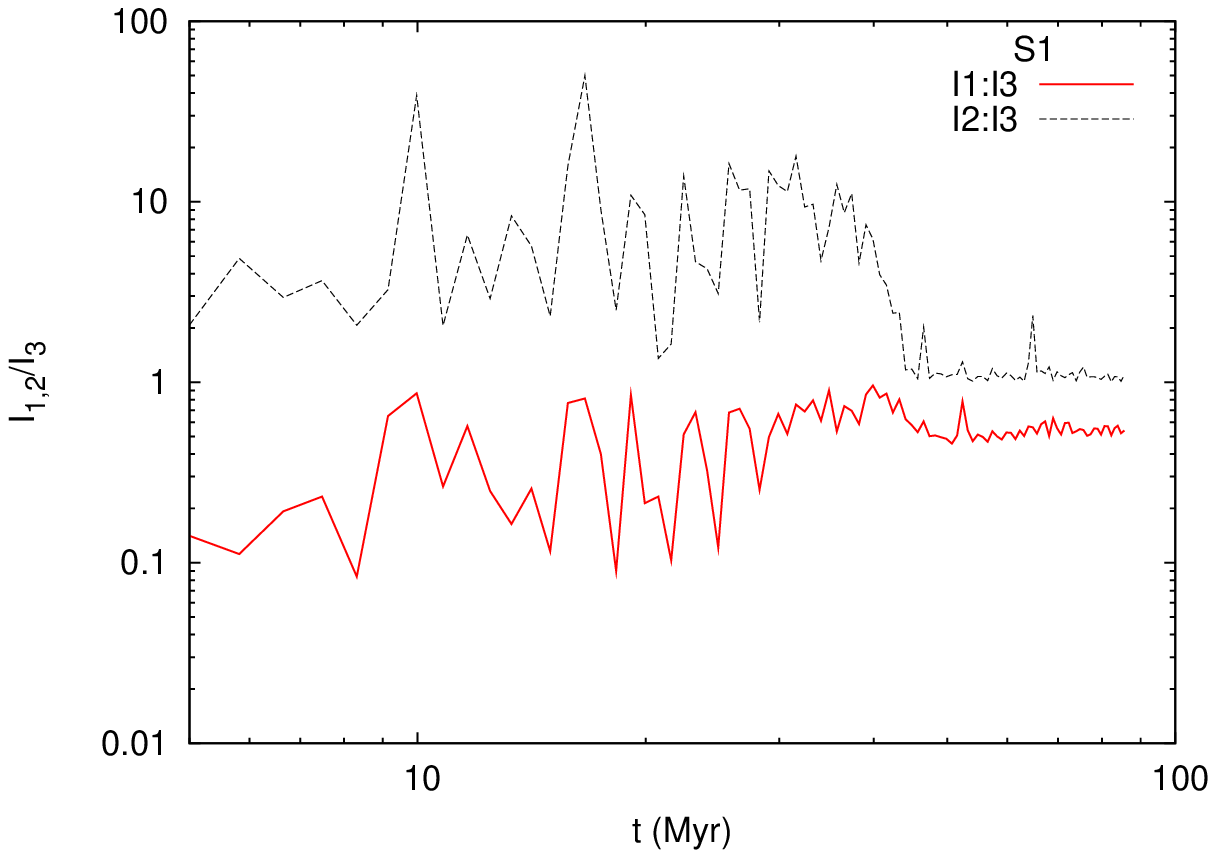}\\
\includegraphics[width=8cm]{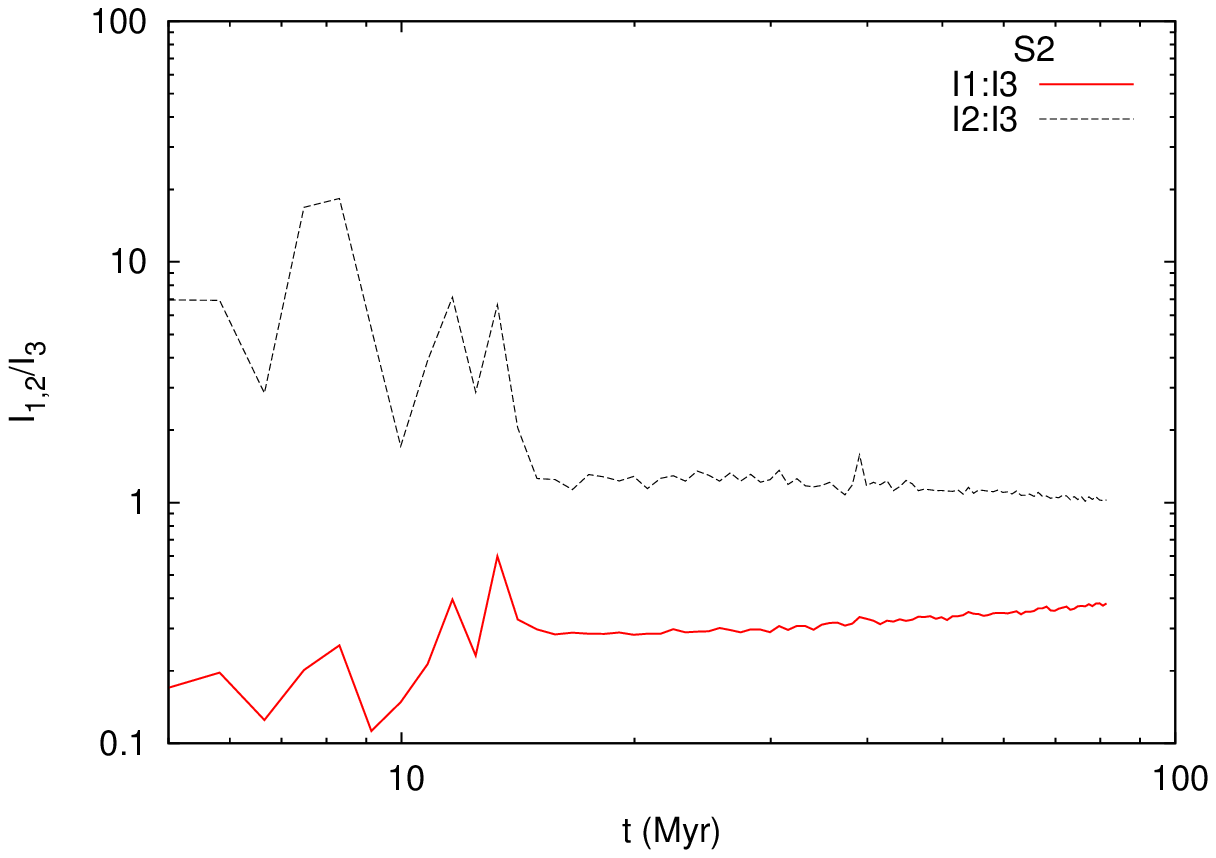}\\
\includegraphics[width=8cm]{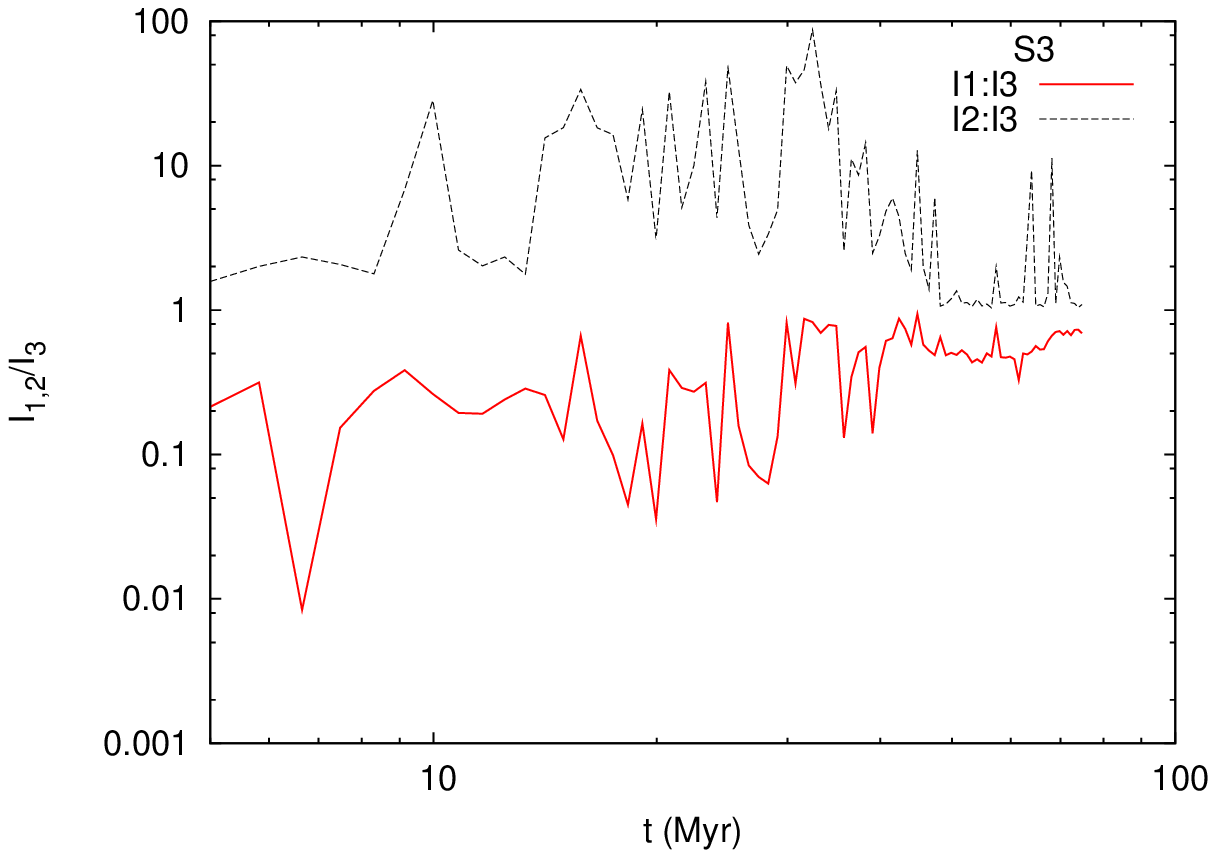}\\
\caption{Principal axes of inertia in our numerical models of the inner 10pc, where the NSC dominates the matter distribution.}
\label{axes}
\end{figure}

In particular, the original disc-like distributions of GCs in model S2 is reflected also in its NSC morphology. Figure \ref{axes} shows the time evolution of its three principal moments of inertia $I_i$, where $I_1$ is the component perpendicular to  orbital plane of clusters in configuration S2.
It is worth noting that all three models are axisymmetric within this preferential plane ($I_2=I_3$), but Model S2 shows an axisymmetic anisotropy with $I_1/I_3\sim 0.3$.

\begin{figure}
\centering 
\includegraphics[width=8cm]{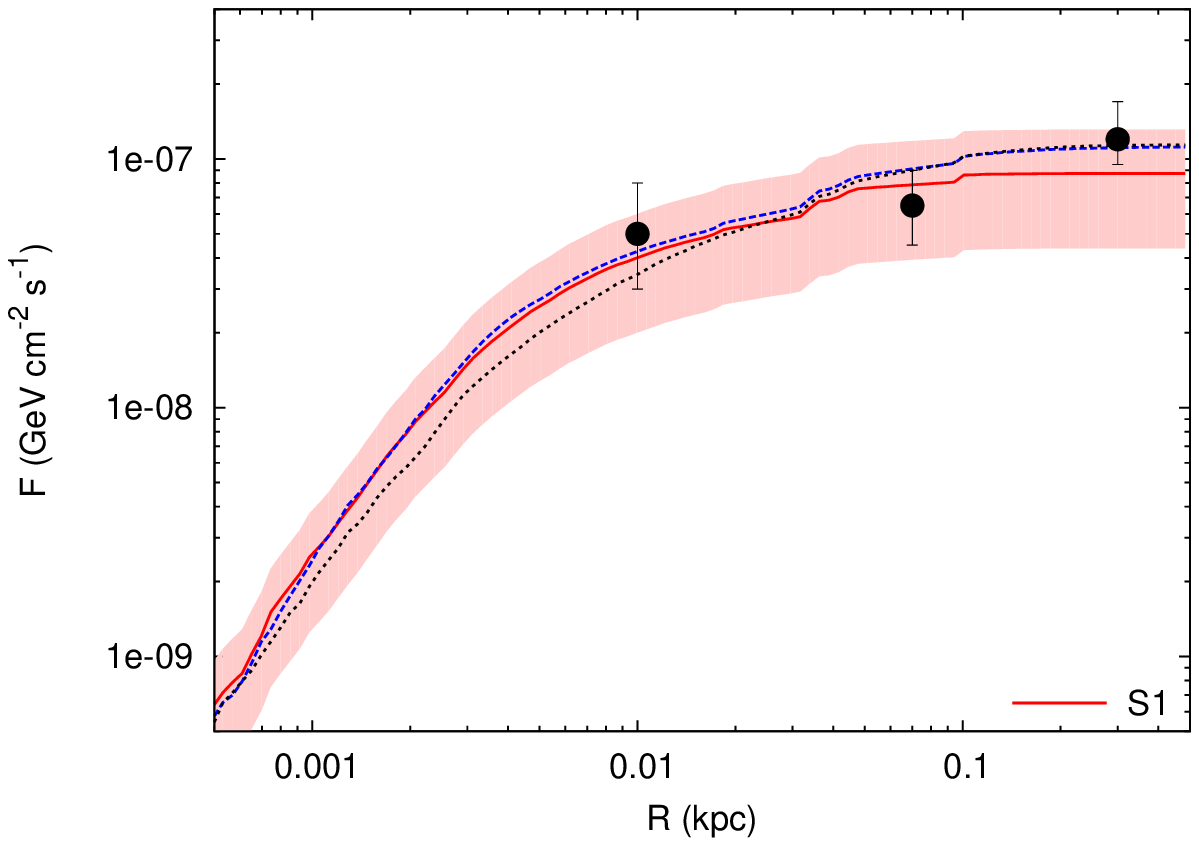}\\
\includegraphics[width=8cm]{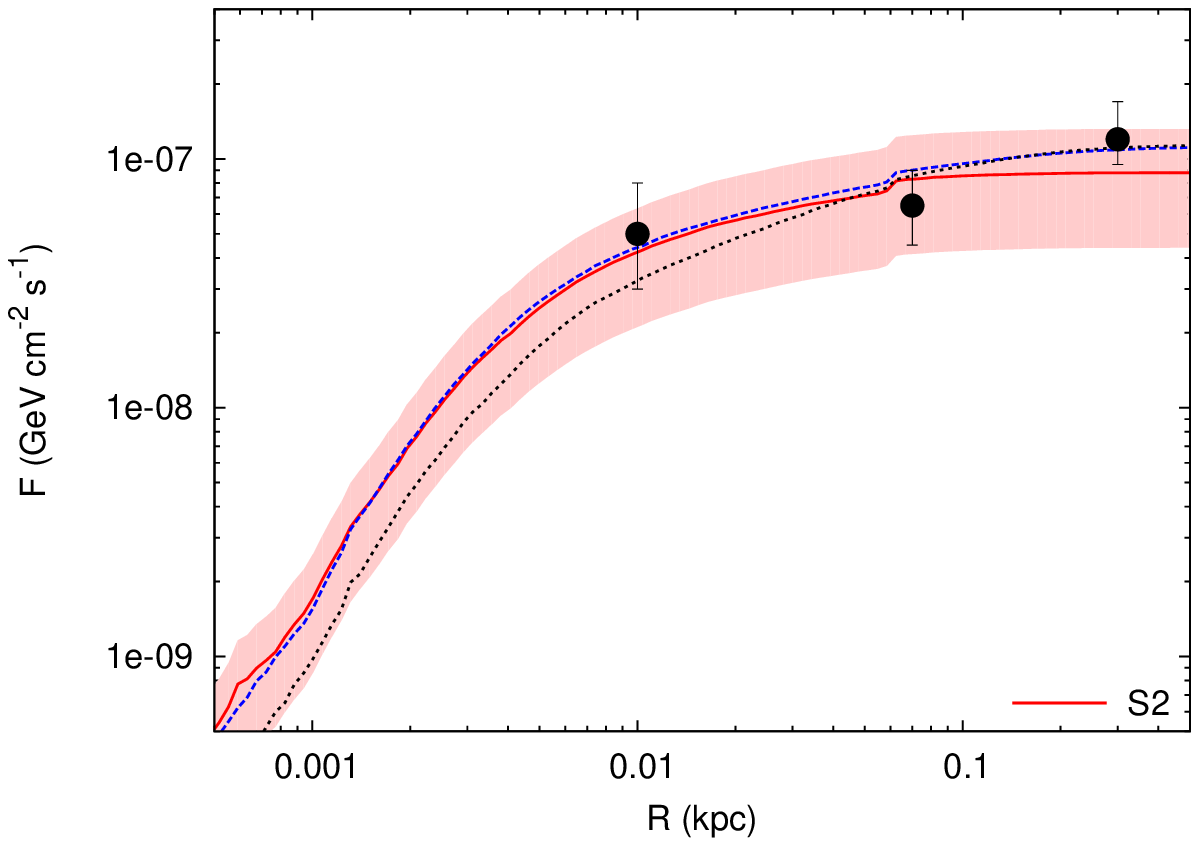}\\
\includegraphics[width=8cm]{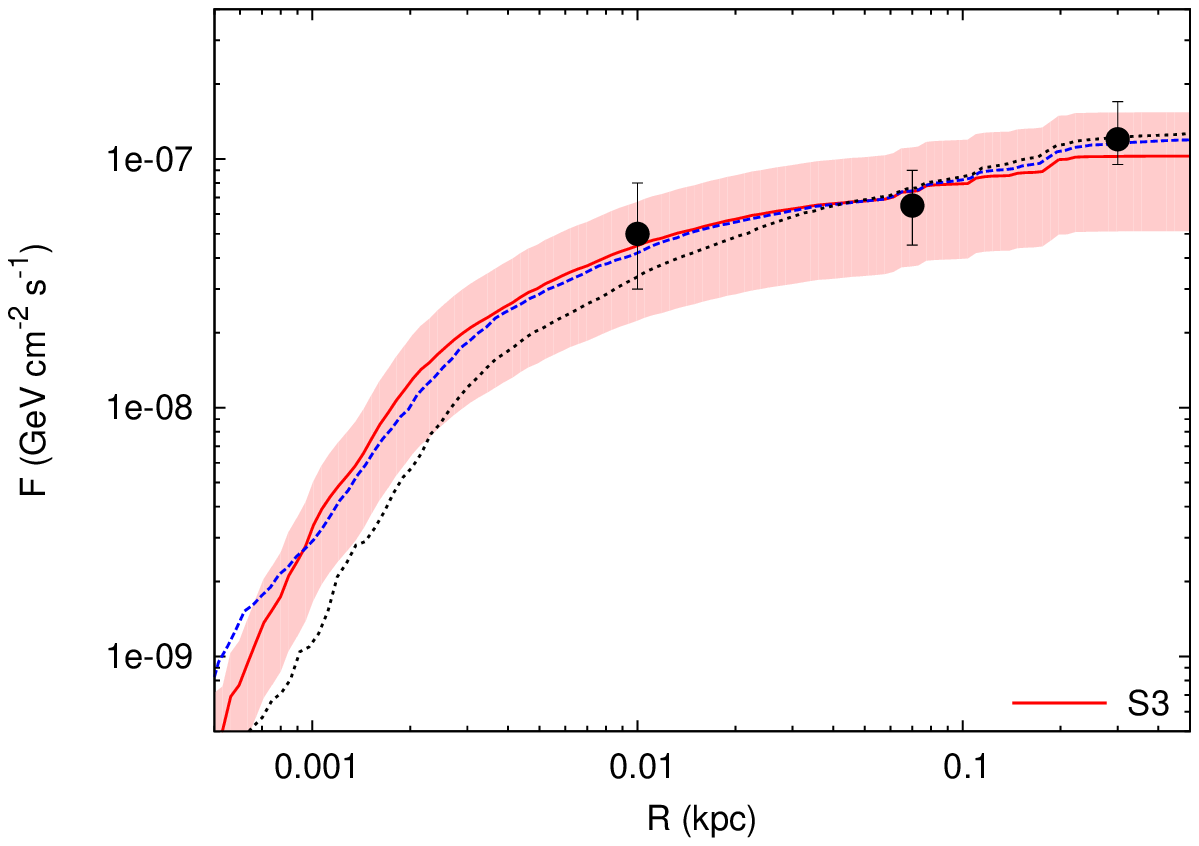}\\
\caption{Cumulative flux distribution calculated from models S1 (top) S2 (centre), S3 (bottom panel), compared with observed $\gamma$ excess \citep[black filled dots][]{brandt15}, as a function of the projected distance to the SMBH. The red shaded region represents the cumulative flux assuming that it comes only from the orbitally segregated clusters. The region width encloses a factor of $50\%$ error in the calculation. The dotted curves are obtained assuming that a fraction of the enclosed galaxy mass contributes to the flux (assuming $\eta = 0.01$)}
\label{map}
\end{figure}

Therefore, gamma-ray imaging of the Galactic centre with a resolution of $6.4$ arcmin carries information on the initial GC population that formed the NSC.
The stellar distribution around the SMBH is expected to be a combination of infalling GC debris and the Galactic background.
Similarly, the gamma-ray flux is expected to have a contribution from these two channels hinting at the relative fraction of ``dry'' and ``wet'' origins of the NSC. To examine these possibilities, we calculate the cumulative flux, $F$, at $2$ Gev similar to that in Figure~\ref{map2gev} above but for all three initial conditions S1, S2, S3 for different values of $\eta_{\rm MSP}$. In Figure \ref{map}, we compare the results with observational estimates by \citet{Abazajian14} (see also Figure 1 in \citealt{brandt15} for a comparison with the modelled cumulative flux).

The results in our three models are quite similar outside of the NSC ($r>10$ pc), while they exhibit interesting differences in the inner few pc, as shown in Figure \ref{map}. The high end of the distribution is compatible with a small contribution coming from the GC, with an upper limit of $\eta \simeq 0.01$, while in the inner pc the presence of the SMBH causes a decrease of the flux by a factor $\sim 2$ for model S3 relative to that in model S1. 

\begin{figure}
\centering 
\includegraphics[width=8cm]{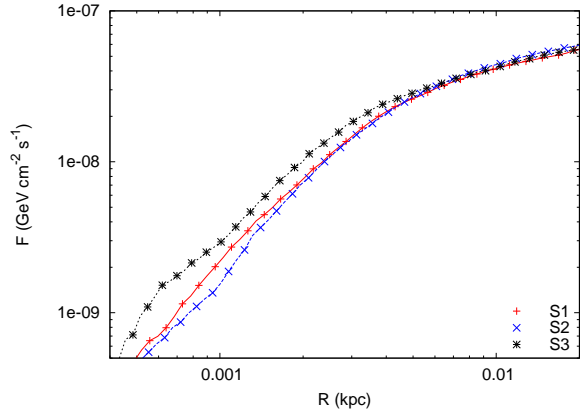}
\caption{Cumulative flux in the three models investigated.}
\label{cumul_comp}
\end{figure}

In Figure \ref{map}, we can identify two important regions: inside $10$ pc the gamma ray flux carries information on both the dry and wet NSC formation pathways, while in the outer region the results are consistent with only a small contribution from the Galactic background. 
The results in the three models investigated look quite similar outside the NSC ($r>10$ pc), while they exhibit interesting differences in the inner few pc.
This is highlighted by Figure \ref{cumul_comp}, which shows the cumulative flux in three models with $\eta \simeq 0.01$.

The shaded region in Figure \ref{map} covers the allowed region assuming a $50\%$ error in our calculations which may be due to i) the uncertainties in the number of MSPs, ii) the level of initial segregation, and iii) the contribution of galactic MSPs.
Regarding the first point, if the GCs reach the Galactic centre before NS form with a velocity dispersion $\sigma = 190$ km s$^{-1}$ \citep{phinney94}, the number of retained neutron stars can increase by a factor up to $\sim 1.5$ with respect to our previous calculation, due to the NSC deeper potential well. Regarding the second point, in deriving the cumulative flux we assumed that all the GCs have the same level of segregation. Accounting for different segregation status may decrease the flux, as expected comparing unsegregated (MSPf) and fully segregated models (MSPa) (see Figure \ref{MSPcom}). Finally, the third point is related to unknown number of MSP formed in the Galactic background, that affects the cumulative flux outside 10 pc.

\begin{figure}
\centering
\includegraphics[width=8cm]{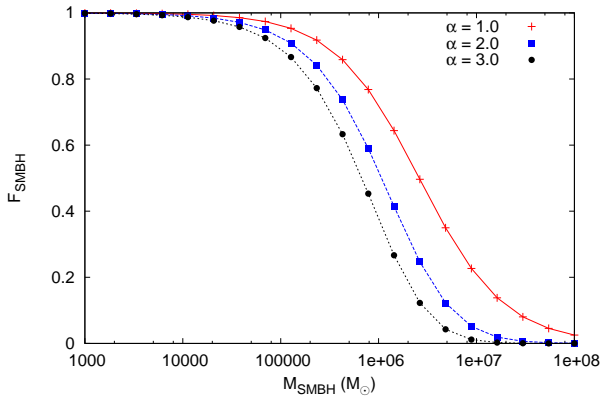}
\caption{Flux emitted from different galactic centres with different SMBHs masses, normalized to the value calculated in absence of an SMBH, as a function of the SMBH mass.}
\label{bhflux}
\end{figure}

We find that the presence of an SMBH in the Galactic centre and the GCs initial conditions cause a noticeable variation of the emission only within 10 pc from the Galactic centre. The $\gamma$ ray flux increases by a factor $\sim 10$ within 1-2 pc in model S3, which do not contain any SMBH. 
The reason for such a difference is related again to the NSC formation process. Indeed, when the SMBH is absent, tidal forces arising from the Galactic centre are significantly smaller. As a consequence, the GCs tidal stripping is less effective in the inner region, allowing for the formation of an NSC characterized by an effective radius smaller than in the other two models (see Table \ref{nsc}).
Therefore, detailed observations of the inner regions of external galaxies that underwent GC-SMBH interactions can potentially help in arguing for the presence of a central SMBH therein. 
This can be particularly interesting in dwarf spheroidal galaxies (dSph), where the relatively low density can prevent the formation of an SMBH seed, depending on the matter distribution in the galaxy \citep{ASCD17}.
For instance, in the simplest approximation in which the emission from the inner 1 pc is connected to the SMBH through a simple power-law, $F_{\rm SMBH} \propto (\frac{M_{\rm SMBH}}{M_0}+1)^{-\alpha}$, where the scaling factor is defined as $M_0=2.6\times 10^6\Ms$, it would be possible to infer the mass of a central SMBH if it exceeds $5\times 10^4\Ms$, almost independently on value of $\alpha$, as shown in Figure \ref{bhflux}.

Moreover, the emission caused by GCs initially orbiting in the same plane should be a few times larger than that caused by GCs moving in the Galactic bulge. Such a difference is quite below the current Fermi resolution, nevertheless these findings can be interesting for next generation of $\gamma$ ray detectors, either space-based, such as the forthcoming ASTROGAM \citep{ASTROGAM}, CALET \citep{CALET} and PANGU \citep{PANGU} space missions, or the CTA telescope \citep{bednarek16} (see \citealt{knodlseder16} for a detailed review on the perspectives of future gamma-ray astronomy).

\subsection{Caveats}

In our analysis we have shown that a population of MSPs and CVs dragged from infalling GCs in the NSC ``dry-merger'' scenario can account for most of the observed hard X-ray and gamma-ray excess flux coming from the innermost region of the Milky Way centre, $r\lesssim 10-100$ pc, while the emission observed outside $\sim 100$ pc is likely due to sources born in-situ. 
On the other hand, our numerical simulations suffer several limitations that are dictated by the current status of numerical modelling of galactic dynamics.

Due to the fact that each of the presented simulations took several months to be completed, the limited number of models provided does not allow us to investigate the role of the GCs' mass function or to determine the impact of a GCs' initial radial distribution significantly different from that of the background Galaxy. As shown in Section \ref{scale}, the resulting NSCs in two models S1 and S2 are both consistent with previous results and observations of the Milky Way NSC, thus suggesting that these assumptions have a minor impact on the NSC final properties.

The mass of each particle in our model is $\sim 70\Ms$, much too high to allow a reliable description of single star dynamics. The ongoing rapid advance in GPU architectures and dedicated programming can eventually boost the level of resolution achievable, and can lead in the near future to lower particle mass values, closer to reality. Numerical codes implementing stellar evolution and strong encounters already exist \citep{aarseth03,spurzem01,petts15,AS16}, but they are still limited to a relatively low number of particles, allowing to model star clusters rather than galactic nuclei \citep{wang16}.

Finally, the processes that regulate the formation rate of MSPs and CVs are still partially unknown, leading to uncertainties in the their predicted numbers in star clusters and in the Galactic centre.
Detailed information on the distribution of MSPs and CVs inside the Milky Way NSC can potentially offer clues to discern whether the Galactic nuclear cluster has a wet or dry origin. These can be combined with further observations of sources compatible with the dry-merger scenario, such as the RRLyrae \citep{min16,dong17}, to further test our conclusions.

\section{Conclusion}
\label{end}

In this paper we investigated possible links between the NSC's origin of the Galaxy and the intense flux observed in the gamma-ray and hard X-ray bands by the Fermi and NuSTAR satellite.
Using state-of-the-art numerical direct $N$-body simulations, we modelled the NSC dynamical formation process by orbital decay and merger of massive star clusters. We investigated the possible configurations of MSPs and CVs in the newly born NSC, delivered in the Galactic centre by the infalling clusters. 

Our main results are summarized as follows.
\begin{itemize}
\item We showed that the dry-merger scenario of GCs provides a suitable mechanism for the deposition of a large number of MSPs and CVs in the Galactic centre. The predicted numbers of MSPs and CVs (particularly IPs) are consistent with the gamma-ray and X-ray observations, assuming that they formed in dense star clusters that underwent orbital decay. 
\item  We found that GCs can deliver up to $\sim 24,000$ BHs to the NSC at the Galactic center. This population is added to the BHs that formed in the NSC.
\item Regarding the gamma-ray emission, our results suggest that nearly $80\%$ of the flux emitted from the inner $\sim 100-150$ pc can be ascribed to MSPs well segregated in their parent stellar clusters, while the remaining $20\%$ can be associated with sources formed in-situ, which dominate the gamma-ray profile outside 20 pc. 
These results are mostly independent of the clusters' initial orbital properties. 
\item The best agreement with the observed gamma-ray emission is achieved assuming that the MSPs' progenitors populate their host clusters' core during the NSC assembly. An originally unsegregated MSP population leads to a final distribution that produces a weaker emission than observed in the range 1--10 pc.
\item The CV number density inferred from our simulations is consistent with observational estimates, while their spatial density profile depends strongly on the level of initial mass segregation in their host clusters. If CVs were initially unsegregated in their parent clusters, their density profile after NSC formation would be flat, while if the whole CVs population is concentrated within the host cluster radius after NSC formation, we get a final power-law density profile, with slope $\sim 0.32$. Therefore, detailed observations of CVs in the Milky Way innermost regions may test these predictions, and shed light on the initial properties of star clusters in the Galactic bulge.
\item The X-ray surface brightness profile inferred from our simulations of the CV population agrees with observations. The best agreement is achieved assuming that $\sim 25\%$ of the CVs in the Galactic centre come from orbitally decayed star clusters, while the remaining X-ray flux is emitted by CVs formed in-situ. The CVs that originated in star clusters (dry merger channel) dominate the emission in the inner 20 pc, while the locally formed sources dominate outside of the NSC. Hence, while the central X-ray emission is mostly due to CVs transported in infalling GCs, the emission coming from outside the NSC is mostly due to a CVs formed in-situ. The difference between in-situ CVs and MSPs is directly related to the number of sources per unit mass, that is much larger for CVs.
\item The star clusters initial orbital parameters determine the morphology of the X-ray and gamma-ray fluxes, but poorly affect the observed cumulative flux distribution in these bands. Star clusters initially distributed accordingly to the Galactic background lead to a triaxial NSC, while a more disky structure forms when the clusters move on co-planar orbits.
\item The central SMBH affects the gamma-ray emission in the inner 10 pc, a limit well below the current FERMI resolution. The absence of a central SMBH leads to the formation of a denser NSC characterised by a flux 5--10 times larger in the inner 1--2 pc than in galaxies with SMBHs. This has interesting implications for the mass range of dwarf galaxies, where the formation of an SMBH may be more difficult due to the lower densities. 
\end{itemize}

\section*{Acknowledgement}

MAS acknowledges the Sonderforschungsbereich SFB 881 "The Milky Way System" (subproject Z2) of the German Research Foundation (DFG) for the financial support provided. The work was done in the footsteps of the  ``MEGaN project: modelling the evolution of galactic nuclei'', funded by the University of Rome Sapienza through the grant 52/2015. 
This project has received funding from the European Research Council (ERC) under the European Union's Horizon 2020 research and innovation programme ERC-2014-STG under grant agreement No 638435 (GalNUC) and from the Hungarian National Research, Development, and Innovation Office grant NKFIH KH-125675 (to B.K.). This work was performed in part at the Aspen Center for Physics, which is supported by National Science Foundation grant PHY-1607761.

\bibliographystyle{mn2e}
\bibliography{ASetal2015}

\end{document}